\documentclass[twocolumn,prl,amsmath,amssymb,showpacs,superscriptaddress,floatfix]{revtex4}
\usepackage{graphicx}
\usepackage{bm}

\usepackage{epsf}
\usepackage[usenames]{color}
\begin{document}
\title{Effect of a potential softness on the solid-liquid transition in a two-dimensional core-softened potential system}

\author{D.E. Dudalov}
\affiliation{ Institute for High Pressure Physics RAS, 142190
Kaluzhskoe shosse, 14, Troitsk, Moscow, Russia}

\author{Yu.D. Fomin}
\affiliation{ Institute for High Pressure Physics RAS, 142190
Kaluzhskoe shosse, 14, Troitsk, Moscow, Russia}
\affiliation{Moscow Institute of Physics and Technology, 141700
Moscow, Russia}

\author{E.N. Tsiok}
\affiliation{ Institute for High Pressure Physics RAS, 142190
Kaluzhskoe shosse, 14, Troitsk, Moscow, Russia}

\author{V.N. Ryzhov}
\affiliation{ Institute for High Pressure Physics RAS, 142190
Kaluzhskoe shosse, 14, Troitsk, Moscow, Russia}
\affiliation{Moscow Institute of Physics and Technology, 141700
Moscow, Russia}

\date{\today}

\begin{abstract}
In the present paper, using a molecular dynamics simulation, we
study a nature of melting of a two-dimensional ($2D$) system of
classical particles interacting through a purely repulsive
isotropic core-softened potential which is used for the
qualitative description of the anomalous behavior of water and
some other liquids. We show that the melting scenario drastically
depends on the potential softness and changes with increasing the
width of the smooth repulsive shoulder. While at small width of
the repulsive shoulder the melting transition exhibits what
appears to be weakly first-order behavior, at larger values of the
width a reentrant-melting transition occurs upon compression for
not too high pressures, and in the low density part of the $2D$
phase diagram melting is a continuous two-stage transition, with
an intermediate hexatic phase in accordance with the
Kosterlitz-Thouless-Halperin-Nelson-Young (KTHNY) scenario. On the
other hand, at high density part of the phase diagram one
first-order transition takes place. These results may be  useful
for the qualitative understanding the behavior of water confined
between two hydrophobic plates.
\end{abstract}


\maketitle

\section{Introduction}

Water plays an important role in many natural processes where it
is confined or at contact with substrates. Examples can be found
in different fields of geology, biology, chemical engineering
because water can be confined in rocks, in biological cells, at
contact with surfaces of proteins, in biological membranes, etc
\cite{water1,water2,water3,water4}. In three dimensions the
qualitative behavior of water, including the waterlike anomalies,
can be described using the core-softened potentials with two
length scales
\cite{wejcp,wepre,we_inv,we2013,RCR,we2013-2,buld2009,fr1,fr2,fr3,buld2011,barb2006,barb2009,prest2,prest3,scala,barbosa,
buld2d,krott,krott1,netz}). The liquids with these potentials
demonstrate anomalous behavior in some regions of thermodynamic
parameters: their phase diagrams have regions where a thermal
expansion coefficient is negative (density anomaly),
self-diffusivity increases upon compression (diffusion anomaly),
and the structural order of the system decreases with increasing
pressure (structural anomaly). This kind of behavior is
characteristic of a number of real fluids, the most common and
well known example is water. Later on it was discovered that many
other substances also demonstrate similar behavior. Some typical
examples are silica, silicon, phosphorus, etc. A lot of different
core-softened potentials were introduced (see, for example,
reviews \cite{buld2009,fr1}). However, it should be noted that in
general the existence of two length scales is not compulsory to
induce the occurrence of the anomalies. For example, for the
models studied in Ref. \cite{prest2,prest3} it was shown that the
existence of two distinct repulsive length scales is not a
necessary condition for the occurrence of anomalous phase
behavior.

In general, the observation, that confined fluids microscopically
relax and flow with different characteristic time scales than bulk
liquids is hardly surprising. Confining boundaries bias the
spatial distribution of the constituent molecules and the ways by
which those molecules can dynamically rearrange. These effects
play important roles in the thermodynamics of the confined systems
and influence the topology of the phase diagram. The general
motivation for the study of different confined fluids follows from
the observation that there are many real physical and biological
phenomena and processes that depend on the properties of such
systems and play an important role in the different fields of
modern technology such as fabrication of nanomaterials,
nanotribology, adhesion, and nanotechnology \cite{rev1,rice}.

It is well known that the nature of spatial ordering of a
molecular system depends on the dimensionality of the space to
which it is confined. As it was shown by Mermin \cite{mermin}, in
$2D$ the long-range crystalline order is destroyed by the thermal
fluctuations and has the  quasi-long-range character. However, the
real long range order does exist in this case - this is the  order
in orientations of bonds between the particle and its nearest
neighbor (bond-orientational order).

During the several decades, the controversy about the nature of
the $2D$ melting transition is lasting. Now it is widely accepted
the theory by Halperin, Nelson, and Young
\cite{halpnel1,halpnel2,halpnel3}, based on the
Kosterlitz-Thouless ideas \cite{kosthoul73} (KTHNY theory). In
accordance with this theory, the scenario of two-dimensional
melting is fundamentally different from the melting of
three-dimensional systems. In $2D$ case the transition between a
crystal and an isotropic liquid can occur by means of two
continuous transitions which correspond to dissociation of bound
dislocation and disclination pairs, respectively. Dissociation of
the dislocation pairs at some temperature $T_m$ destroys the
quasi-long-range translational order. The resulting phase has the
short-range translational order, but the quasi-long-range bond
orientational order. This phase is called the hexatic phase. The
isotropic liquid appears as a result of dissociation of the
disclination pairs at some temperature $T_i$.

The KTHNY theory have got an experimental support from the
experiments with electrons on helium \cite{gram} and the colloidal
model system with repulsive magnetic dipole-dipole interaction
\cite{keim1,zanh,keim2,keim3,keim4}. In the case of colloidal
suspension, confined in a channel, reducing the channel size
changes the system behavior from three-dimensional to
two-dimensional, and colloidal monolayer displays continuous $2D$
melting \cite{col1,col2,col3,col4}.

At the same time, a conventional first-order transition between a
two-dimensional solid and an isotropic liquid can also occur
(\cite{chui83,klein2,rto1,rto2,RT1,RT2,ryzhovTMP,ryzhovJETP}). For
example, as it was shown in \cite{ryzhovTMP,ryzhovJETP}, at low
disclination core energy system can melt through one first-order
transition as a result of the dissociation of the disclination
quadrupoles.

Despite numerous experimental and simulation studies, the
situation is controversial - KTHNY theory seems universal and
independent on the pair potential of the system, however, the
systems with very short range or hard core potentials demonstrates
the coexisting phases (weak first-order transition), while the
melting scenarios for the soft repulsive particles favor the KTHNY
theory
\cite{rice,prest2,DF,LL,prest1,dfrt1,dfrt2,dfrt3,rice1,rice2,strandburg92,binderPRB,
mak,jaster2,jaster3,andersen,andersen1,hfo1,hfo2,binder,ss1,ss2,ss3,LJ,gribova}.

Even for simplest potential systems, including hard and soft disks
or Lennard-Jones potentials, despite the tendency to the weak
first-order melting transition, the simulation results are
conflicting
\cite{strandburg92,binderPRB,mak,jaster2,jaster3,andersen,andersen1,hfo1,hfo2,binder,ss1,ss2,ss3,LJ}.
The "toy" model for the investigation of the melting transition in
$2D$, hard-disk system, have been studied in a large number of
computer simulations without unambiguous conclusion about the
melting scenario \cite{binderPRB,mak,jaster2,jaster3,binder}. It
was even suggested an unexpected possibility of the first-order
hexatic-isotropic liquid transition \cite{hfo1,hfo2}.

In general, one can conclude that the melting transition strongly
depends on the pair interaction in the system. For instance, as
shown by Bladon and Frenkel \cite{DF} in the framework of the
computer simulations, a $2D$ solid of particles with short-range
attraction can be unstable to dislocation unbinding in a region
that is clearly thermodynamically stable with respect to the
isotropic fluid. The system supports an isostructural solid--solid
transition. In the vicinity of the critical temperature for this
isostructural transition, fluctuations can induce formation of a
hexatic phase. As the range of attraction part of the potential
grows, the hexatic phase region becomes larger and moves toward
the melting line.

In Refs. \cite{RT1,RT2} the density functional calculations have
shown that the $2D$ square-well system can demonstrate both
first-order and continuous melting phase transitions depending on
the width of the attractive well.

Recently, Prestipino et al \cite{prest1} presented a Monte Carlo
simulation study of the phase behavior of two-dimensional
classical particles repelling each other through an isotropic
Gaussian potential \cite{gauss1,prest4,gauss2,gauss3, we_gauss}.
They have shown, that as in the analogous three-dimensional case,
a reentrant-melting transition occurs along with a spectrum of
waterlike anomalies in the fluid phase. However, in $2D$ melting
is a continuous two-stage transition, with a narrow intermediate
hexatic phase, in agreement with the KTHNY scenario.

In Ref. \cite{prest2}, Prestipino, Saija, Giaquinta considered the
behavior of the system with the extremely soft Yoshida and
Kamakura potential \cite{YK}. It was shown that in the model there
exist three different crystal phases, one of them with square
symmetry and the other two triangular. The most interesting fact
is that the triangular solids melt into a hexatic fluid, while the
square solid is directly transformed on heating into an isotropic
fluid through a first-order transition. A whole spectrum of
waterlike anomalies also exists for this model potential.

In our recent papers \cite{dfrt1,dfrt2,dfrt3} we presented a
computer simulation study of the phase diagram and anomalous
behavior of two-dimensional core-softened potential system
\cite{wejcp,wepre,we_inv,we2013,RCR,we2013-2} with the soft core
$\sigma_1=1.35$ (see Eq. (\ref{3})). As in Ref. \cite{prest2}, we
found three different crystal phases, one of them with square
symmetry and the other two triangular. In the low density part of
the $2D$ phase diagram, melting of the triangular phase is a
continuous two-stage transition, with an intermediate hexatic
phase. All available evidences support the KTHNY scenario for this
melting transition. At high density part of the phase diagram
square and triangular phases melt through one first-order
transition.

On the other hand, in \cite{buld2d} the phase diagram was studied
for a square-shoulder square-well potential in two dimensions that
has been previously shown to exhibit liquid anomalies consistent
with a metastable liquid-liquid critical point \cite{scala}. It
was shown that all the melting lines are first order, despite a
small range of metastability.

This paper extends our previous results to the set of systems
which are characterized by the different softness, in order to
study the influence of the softness on the melting scenarios. We
present a simulation study of phase behavior of the core-softened
system, introduced in our previous publications
\cite{wejcp,wepre,we_inv,we2013,RCR,we2013-2,dfrt1,dfrt2,dfrt3},
in two dimensions. The main goal is to compare the phase diagrams
of the systems with the potentials of the increasing softness and
to reveal the influence of the softness of the potential on the
melting scenarios.

The paper is organized as follows: Sec. II presents the system and
methods, Sec. III describes the results and their discussion,  and
finally Sec. IV contains conclusions.

\section{Systems and methods}

\begin{figure}
\begin{center}
\includegraphics[width=8cm]{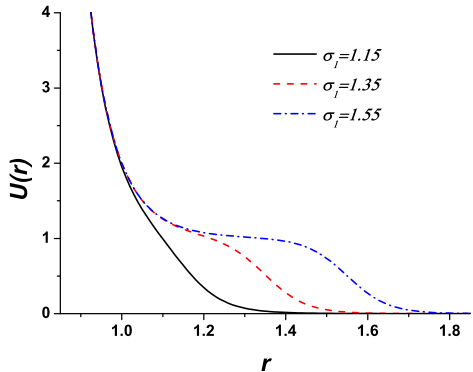}%

\end{center}

\caption{\label{fig:fig1} The potential (\ref{3}) with three
different soft-core diameters:$\sigma_1 = 1.15; 1.35; 1.55$.}
\end{figure}

The system we study in the present simulations is the smooth
repulsive shoulder system (SRSS) introduced in our previous
publications
\cite{wejcp,wepre,we_inv,we2013,RCR,we2013-2,dfrt1,dfrt2,dfrt3}:
\begin{equation}
U(r)=\varepsilon\left(\frac{\sigma}{r}\right)^{n}+\frac{1}{2}\varepsilon\left(1-
\tanh(k_1\{r-\sigma_1\})\right). \label{3}
\end{equation}
where $n = 14$ and $k_1\sigma = 10.0$. $\sigma$ is the hard-core
diameter. We simulate the systems with three different soft-core
diameters: $\sigma_1/\sigma = 1.15; 1.35; 1.55$. (see
Fig.~\ref{fig:fig1}).

In the remainder of this paper we use the dimensionless
quantities, which in $2D$ have the form: $\tilde{{\bf r}}\equiv
{\bf r}/\sigma$, $\tilde{P}\equiv P \sigma^{2}/\varepsilon ,$
$\tilde{V}\equiv V/N \sigma^{2}\equiv 1/\tilde{\rho}, \tilde{T}
\equiv k_BT/\varepsilon, \tilde{\sigma_1}=\sigma_1/\sigma$. Since
we will use only these reduced variables, the tildes will be
omitted.

As it was discussed before \cite{wejcp}, the system with this
potential corresponds to the quasibinary mixture of spheres with
the diameters $\sigma$ and $\sigma_1$.

As it was shown in \cite{wejcp,wepre,we_inv,we2013,RCR,we2013-2},
in $3D$ system of particles interacting through a core-softened
potential (\ref{3}) exhibits a number of unusual features,
including reentrant melting, a maximum on the melting curve,
superfragile glass behavior, and thermodynamic and dynamic
anomalies similar to the ones found in water and silica.

In our previous publications \cite{RT1,RT2} it was shown that one
has to distinguish two temperatures which characterize the melting
transition in two dimensions. The mean-field temperature $T_{MF}$
corresponds to the disappearance of the modulus of the Fourier
transform of the local density $\rho_{\bf G}$ and another one,
$T_m$, is the temperature at which the singular fluctuations of
the phase of the order parameter (free dislocations) appear in the
system. In the case $T_m<T_{MF}$ one can expect that the system
melts through the continuous Kosterlitz-Thouless transition,
$T_{MF}$ being the limit of the metastable solid phase. If
$T_m>T_{MF}$, the system melts via a first-order transition.
$T_{MF}$ can be obtained from the double-tangent construction for
the free energies of liquid and solid phases, but in order to
determine $T_m$ and  $T_i$ and conclude whether the melting occurs
through the KTHNY scenario, the additional analysis is necessary.

In order to distinguish between the first-order and continuous
melting, we also used the criteria described in Ref.
\cite{binder}. We calculated the pressure $P$ versus density
$\rho$ along the isotherms, the bond orientational $\psi_n, n=4,6$
and translational $\psi_T$ order parameters (OPs), which
characterize the overall translational and orientational order,
and the bond orientational correlation function $G_n(r), n=4,6$.
In the case of the first-order phase transition the isotherms
demonstrate the Van der Waals loops, and OP $\psi_n$ has almost
linear behavior in the supposed two-phase region as a function of
density. According to the KTHNY scenario
\cite{halpnel1,halpnel2,halpnel3,binder}, melting in $2D$ can
occur via two continuous transitions (at densities $\rho_l$ and
$\rho_s$ ) with corresponding pressures $P_l$ and $P_s>P_l$,
rather than via a single first-order transition at coexistence
pressure $P_T$ (where a fluid of density $\rho_l$ and a solid of
density $\rho_s$ will coexist). In this case the isotherm are
smooth in contrast with the Van der Waals loops for a first order
transition \cite{binder}.

We use the molecular dynamics simulations of the system in $NVT$
and $NVE$ ensembles (LAMMPS package \cite{lammps}) with the number
of particles between $3200$ and $102400$. To determine $T_{MF}$ as
a function of density,  we calculate the Helmholtz free energy for
liquid and solid phases and construct a common tangent to them.
Because the potential (\ref{3}) is purely repulsive, there is no
liquid-gas transition in the system. In this case, the Helmholtz
free energy of the liquid can be calculated  by integrating the
equation of state along an isotherm \cite{book_fs}:
\begin{equation}
\frac{F(\rho)-F_{id}(\rho)}{Nk_BT}=\frac{1}{k_BT}\int_{0}^{\rho}\frac{P(\rho')-\rho'
k_BT}{\rho'^2}d\rho'. \label{hfe}
\end{equation}
Free energies of solid phases were calculated in the framework of
the method of coupling to the Einstein crystal \cite{book_fs}.
Double tangent construction gives the lines of the first order
melting transition.

In order to proceed in analysis of the melting scenarios, let us
define the translational order parameter $\psi_T$ (TOP), the
orientational order parameter $\Psi_n$ (OOP), and the
bond-orientational correlation function $G_n(r)$ (OCF) in the
conventional way
\cite{prest2,halpnel1,halpnel2,strandburg88,binder,binderPRB,LJ,prest1}.

TOP is taken in the form
\begin{equation}
\psi_T=\frac{1}{N}\left<\left|\sum_i e^{i{\bf G
r}_i}\right|\right>, \label{psit}
\end{equation}
where ${\bf r}_i$ is the position vector of particle $i$ and {\bf
G} is the first shell reciprocal-lattice vector of the crystal. It
should be noted that $\psi_T$ is nonzero if a solid is oriented
consistently with the length and direction of {\bf G}. In the
simulation, $\psi_T$ is measured on heating of the large enough
crystal where the original crystal orientation is mainly
preserved. Melting of the crystal phase into hexatic phase or
isotropic liquid is determined by the sharp decrease of $\psi_T$.

To study the orientational order and the hexatic phase, let us
define the local order parameter, which measures the $n$-fold
orientational ordering, in the following way:
\begin{equation}
\Psi_n({\bf r_i})=\frac{1}{n(i)}\sum_{j=1}^{n(i)} e^{i
n\theta_{ij}}\label{psi6l},
\end{equation}
where $\theta_{ij}$ is the angle of the bond between particles $i$
and $j$ with respect to a reference axis and the sum over $j$ is
over all $n(i)$ nearest-neighbors of $j$, found from the Voronoi
construction. The global OOP is obtained as an average over all
particles:
\begin{equation}
\psi_n=\frac{1}{N}\left<\left|\sum_i \Psi_n({\bf
r}_i)\right|\right>.\label{psi6}
\end{equation}
It should be noted that $n=6$ corresponds to the triangular solid
and $n=4$ - to square solid. In a perfect triangular solid
$n(i)=6$, $\theta_{ij}=\pi/3$ and $\psi_6=1$.

The bond-orientational correlation function $G_n(r)$ (OCF) is
given by the equation:
\begin{equation}
G_n(r)=\left<\Psi_n({\bf r})\Psi_n^*({\bf 0})\right>, \label{g6}
\end{equation}
where $\Psi_n({\bf r})$ is the local bond-orientational order
parameter (\ref{psi6l}).

Both in the isotropic fluid phase and in the hexatic phase,
$\psi_n\rightarrow 0$ as $L\rightarrow 0$, where $L$ is the linear
size of the system, but the behaviors of $G_n(r)$ are different.
The KTHNY theory predicts an algebraic large-distance decay of the
OCF in the hexatic phase, which should be contrasted with the
exponential asymptotic vanishing of angular correlations in a
normal isotropic fluid:
\begin{eqnarray}
G_n(r)&=&e^{-r/\xi}, r\rightarrow \infty, \rho<\rho_l,
\label{g61}\\
G_n(r)&=&r^{-\eta(T)}, r\rightarrow \infty, \rho_l<\rho<\rho_s
\label{g62}.
\end{eqnarray}
Here $\xi$ is the correlation length of the bond orientational
order, which diverges as $\rho_l$ is approached. Another
prediction of the theory is $\eta=1/4$ at the hexatic-to-normal
isotropic fluid transition point
\cite{halpnel1,halpnel2,halpnel3}.

\section{Results and discussions}

\begin{figure}
\includegraphics[width=8cm]{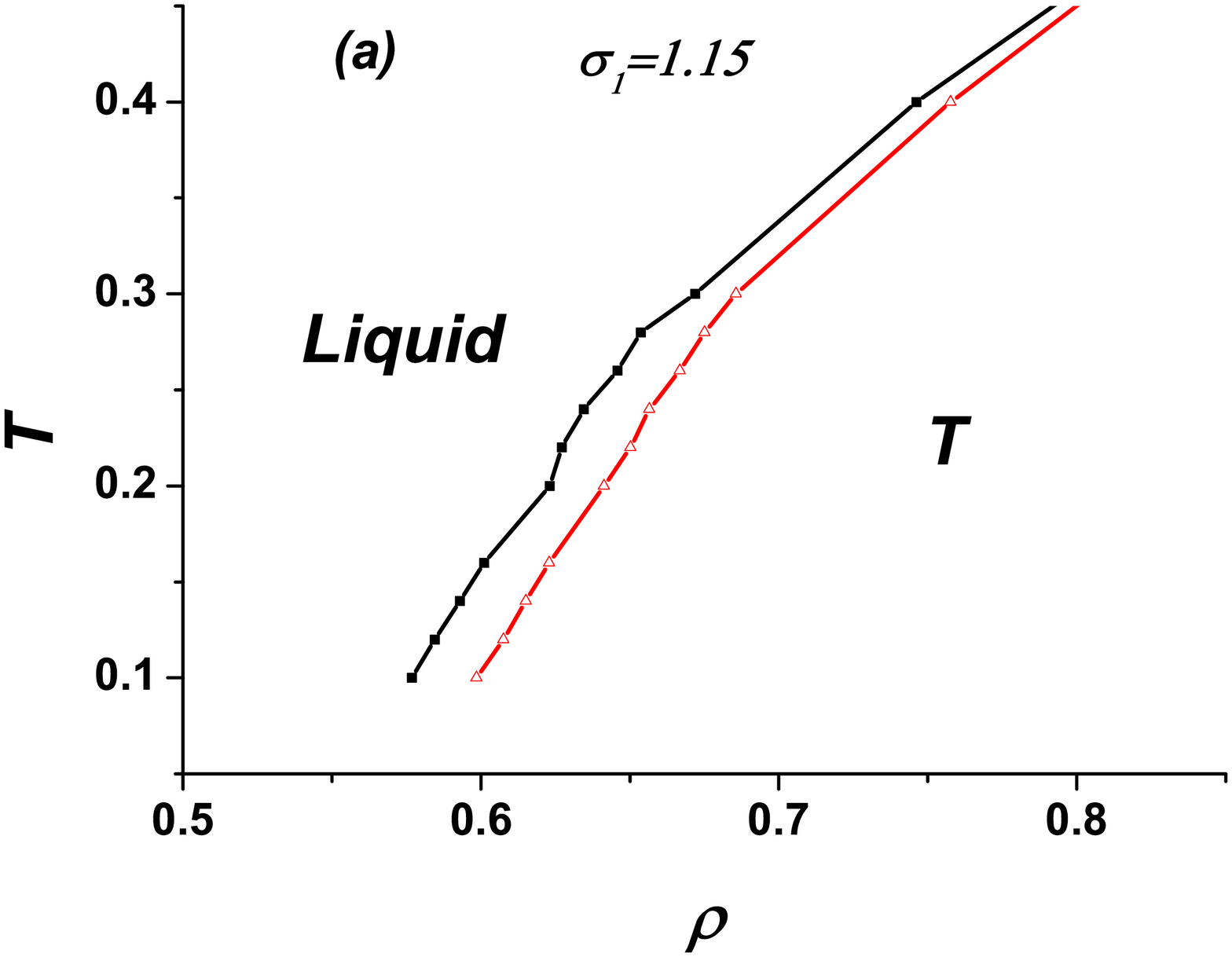}

\includegraphics[width=8cm]{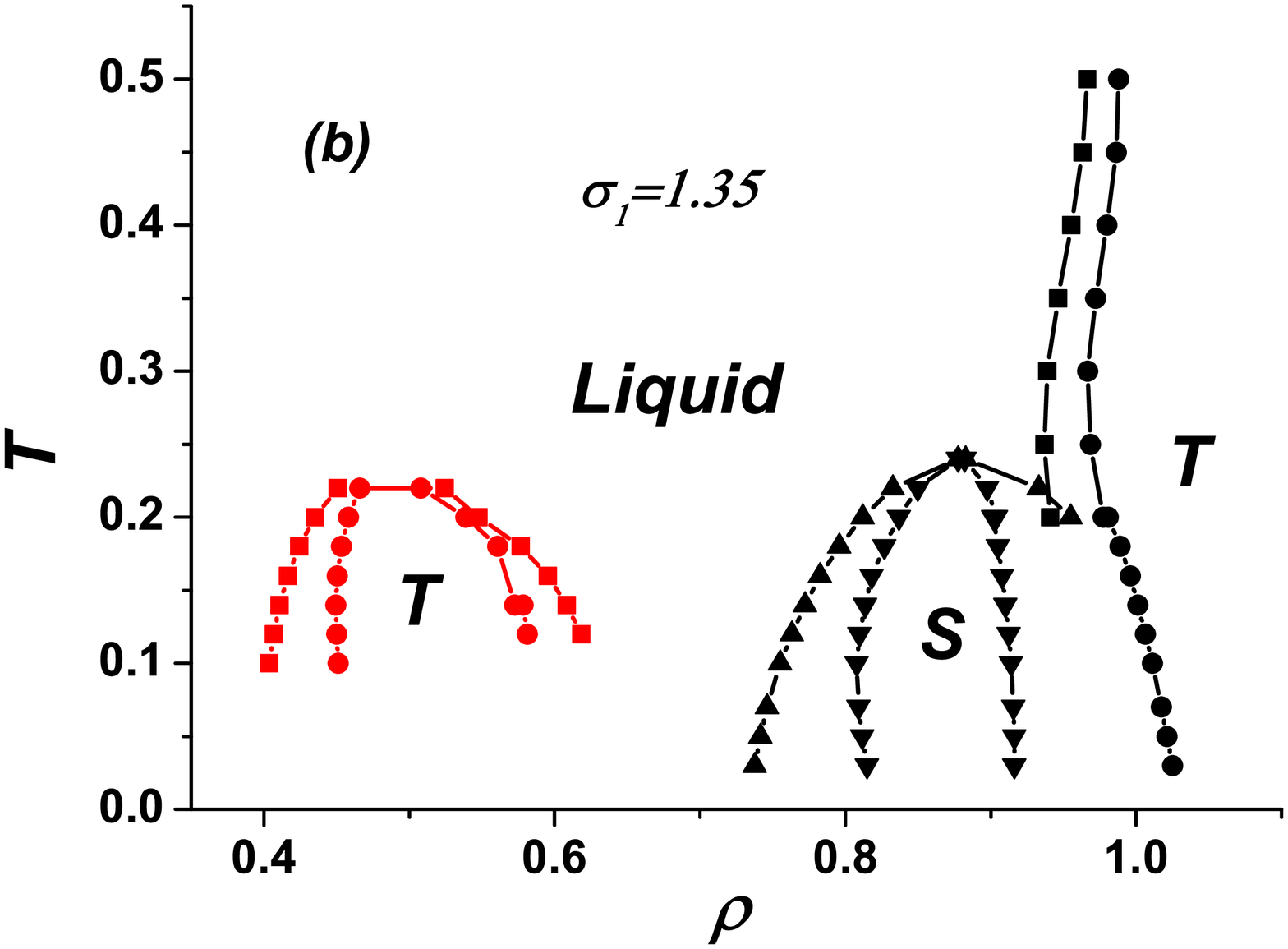}

\includegraphics[width=8cm]{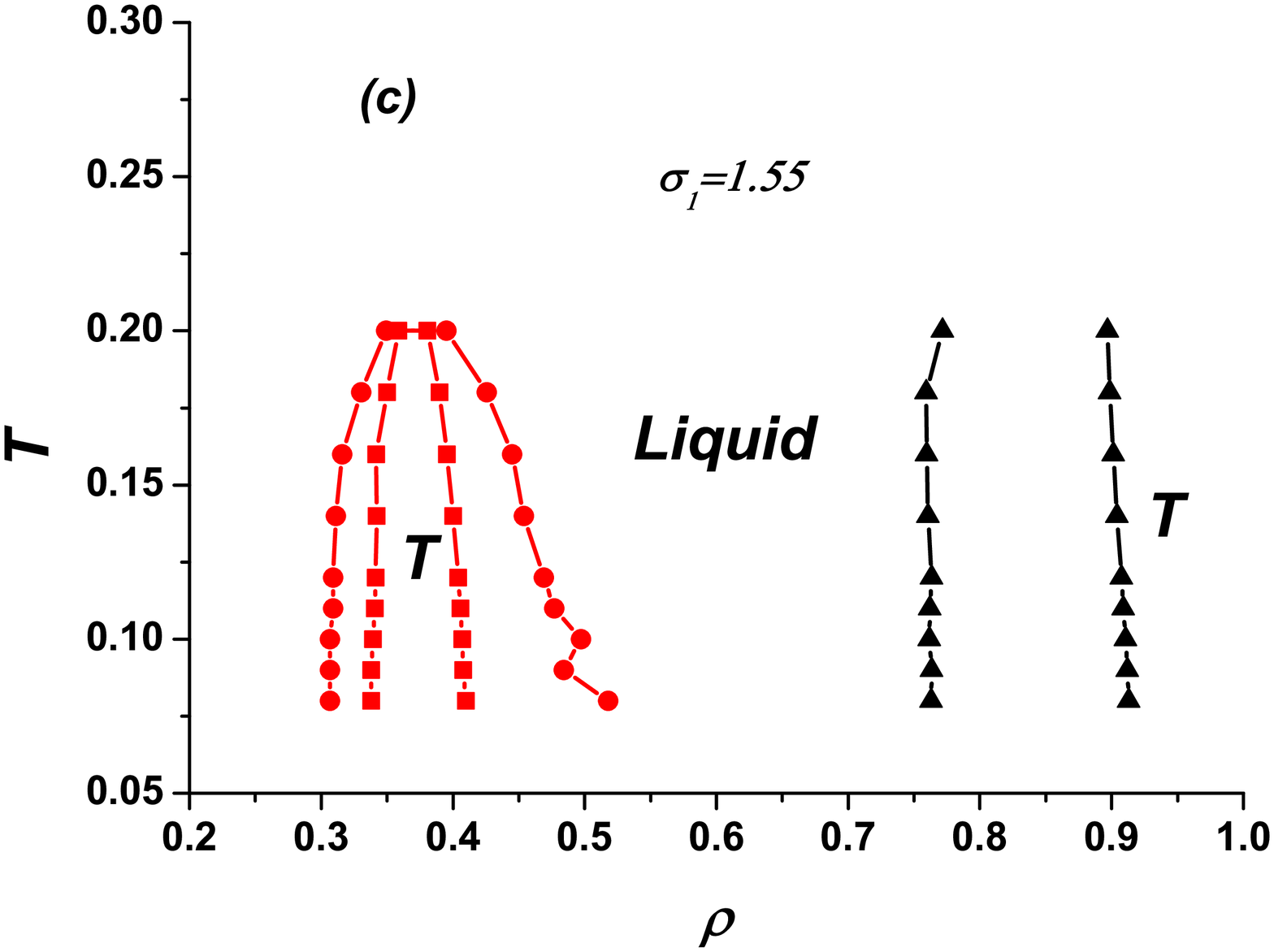}

\caption{\label{fig:fig2} (Color online) (a) Phase diagram of the
$2D$ system with the potential (\ref{3}) for $\sigma_1=1.15$ in
$\rho-T$ plane, where liquid and Triangular (T) phases are shown.
(b) Phase diagram for $\sigma_1=1.35$, where Triangular (T) and
Square (S) phases are shown. (c) Phase diagram of the same system
in the $\rho-T$ plane for $\sigma_1=1.55$, where liquid and
Triangular (T) phases are shown.}
\end{figure}

In Fig.~\ref{fig:fig2} we plot the phase diagrams of $2D$ system
with the potential (\ref{3}) for $\sigma_1=1.15, 1.35, 1.55$ in
$\rho-T$ coordinates, obtained from the double tangent
construction calculations. As it was mentioned above, these phase
diagrams correspond to the thermodynamic limits of the stability
of solid phases and give the first order melting lines. In order
to check the possibility of the continuous melting, the further
analysis is necessary. One can see that for $\sigma_1=1.15$
(Fig.~\ref{fig:fig2}(a)) there are only liquid and triangular
solid phases, and the phase diagram is similar to the ordinary
soft spheres case where the weak first order transition can be
expected \cite{strandburg88,ss1,ss2,ss3}.

For $\sigma_1=1.35$ (Fig.~\ref{fig:fig2}(b)) the phase diagram is
more complex. It was discussed recently \cite{dfrt1,dfrt2,dfrt3}
in detail, and here we recall some main points. There is a clear
maximum in the melting curve at low densities. The phase diagram
consists of two triangular crystal domains (T) corresponding to
close packing of the small and large disks separated by a
structural phase transition and square lattice (S). It is
important to note that there is a region of the phase diagram
where we have not found any stable crystal phase at the
temperatures accessible in our simulations. The results of $3D$
simulations \cite{wejcp,RCR} suggest that a glass transition can
occur in this region. It should be noted, that for this system the
phase diagrams in $2D$ and $3D$ are qualitatively similar
\cite{wejcp,dfrt3}. They consist in two structures which are close
packed in corresponding dimensions: FCC in $3D$ and triangular in
$2D$, between which there are some other phases which depend on
the parameters of the potential and dimensionality. These
structures correspond to the crystalline phases of small spheres
(at high densities) and large spheres (at low densities). The
qualitative shape of the phase diagrams is determined by the
existence of two scales in the potential \cite{wejcp}.

The case $\sigma_1=1.55$ (see Fig.~\ref{fig:fig2}(c)) is
qualitatively similar to the previous phase diagram, however, the
only triangular solid was found. The "gap" between two ordered
parts of the phase diagram is wider, but it seems that in the
temperature range explored in our simulations no other simple
crystal structure can exist. For instance, we found that the
square lattice, which can be supposed to exist in analogy with the
case $\sigma_1=1.35$ (Fig.~\ref{fig:fig2}(b)) is unstable at the
temperatures used in our simulations. It should be noted, that in
$3D$ the similar situation takes place
\cite{wejcp,wepre,we_inv,we2013,RCR,we2013-2}, however, for
$\sigma_1=1.35$ it was shown \cite{prest5}, that at $T=0$ there
are rather complex crystal structures in the density range, where
the "gap" in the phase diagram exists. Unfortunately, we could not
find similar structures in $2D$.

\begin{figure}
\includegraphics[width=8cm]{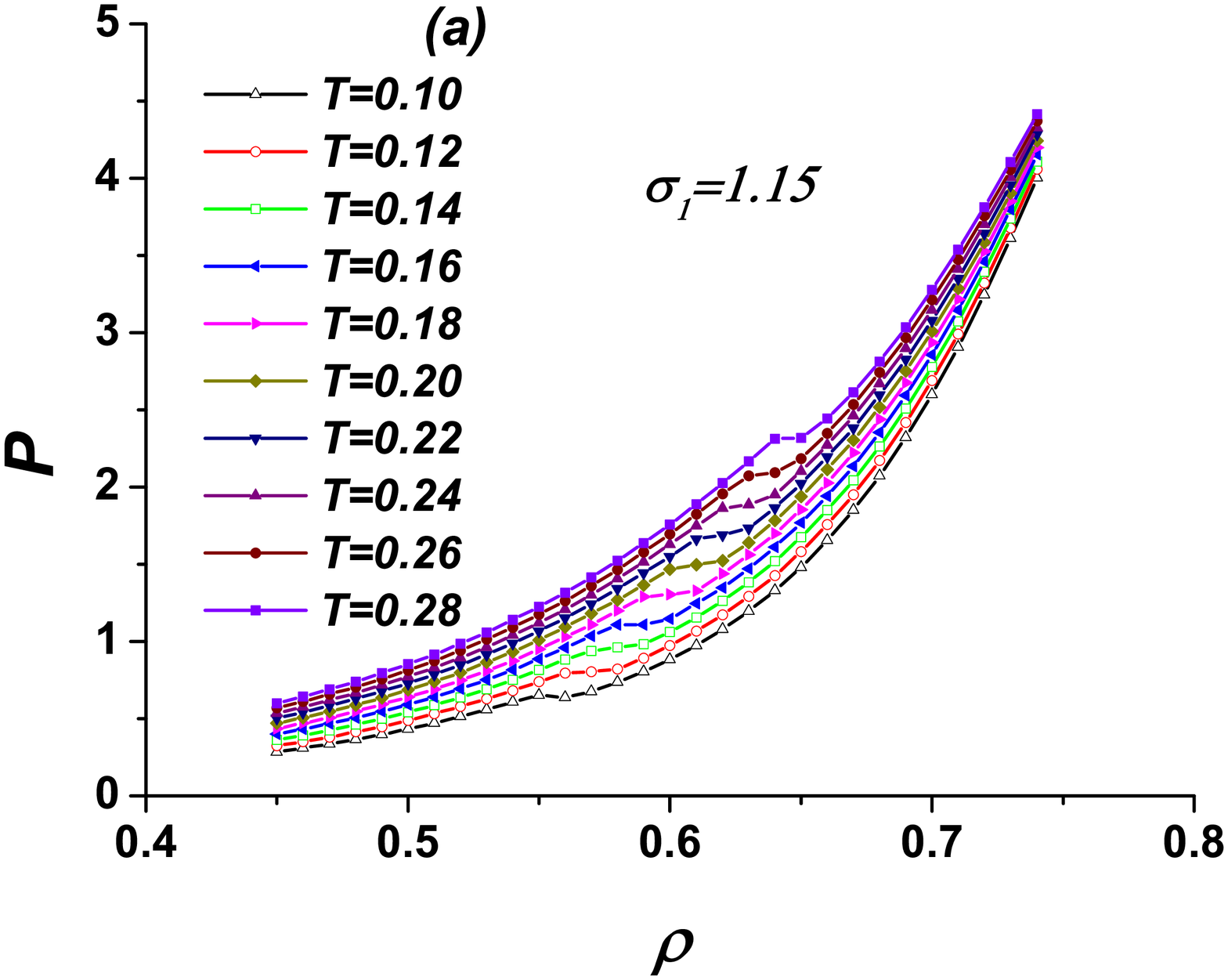}

\includegraphics[width=8cm]{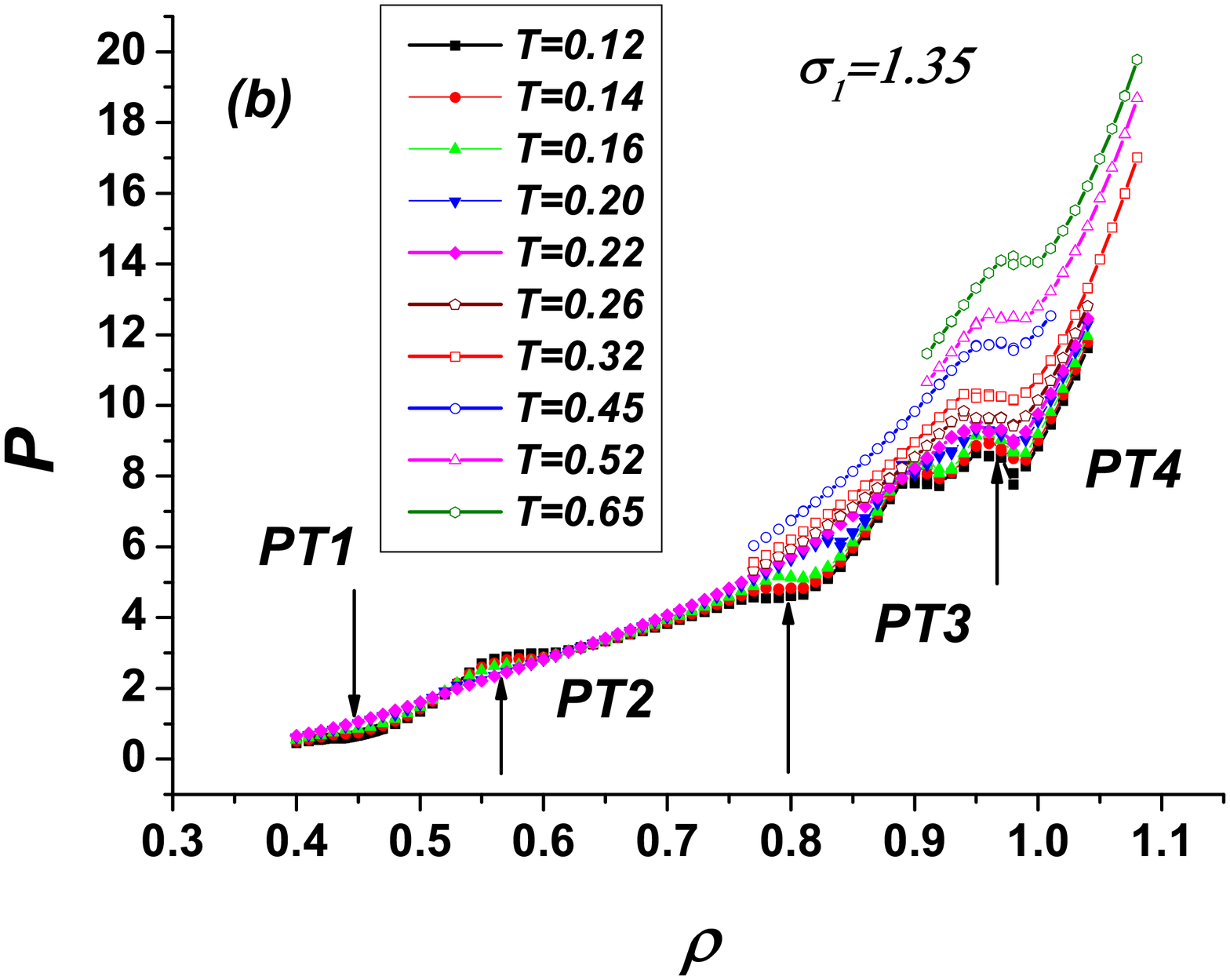}

\includegraphics[width=8cm]{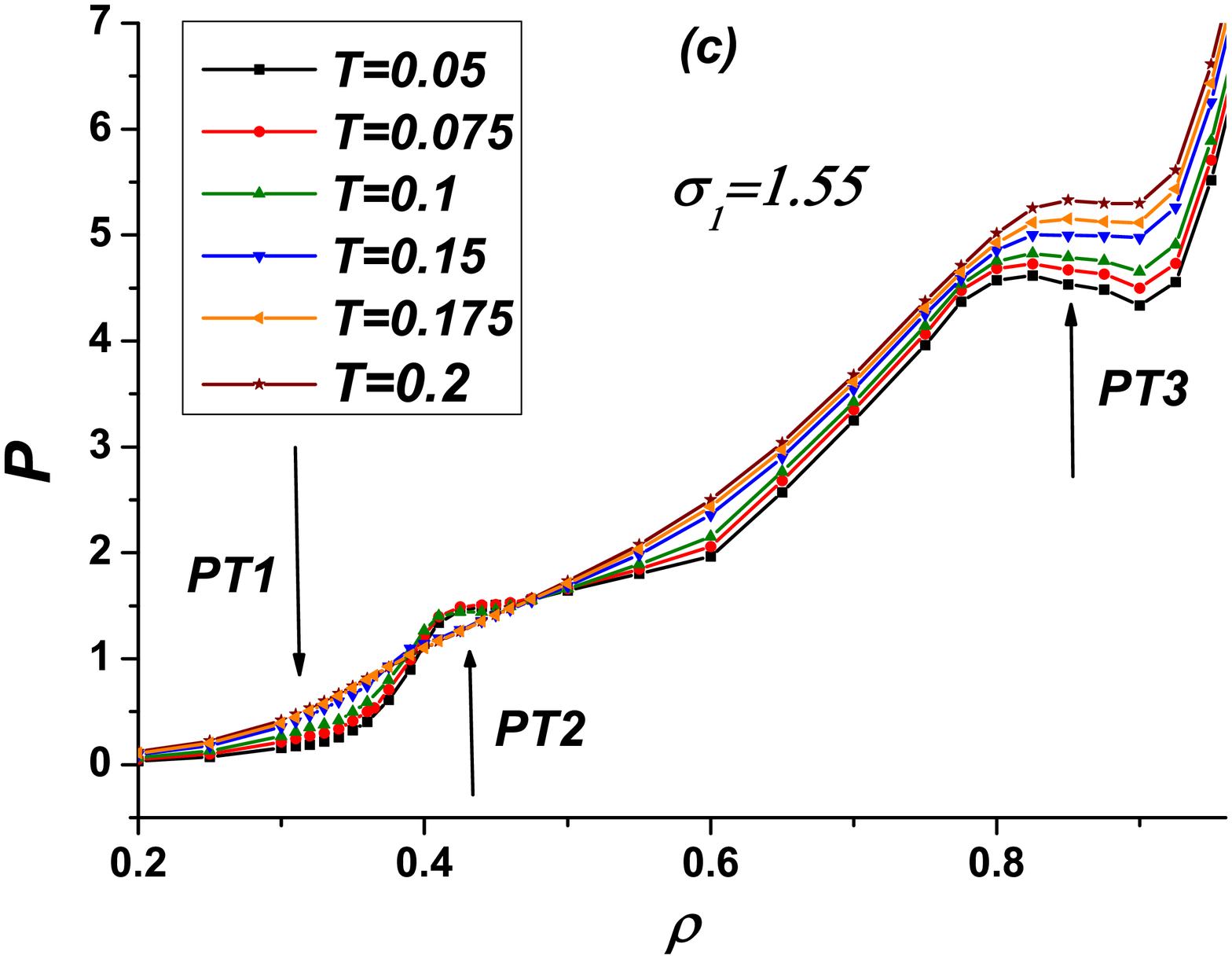}

\caption{\label{fig:fig3} (Color online) (a) The set of isotherms
for $\sigma_1=1.15$.  (b) The set of isotherms for
$\sigma_1=1.35$. The arrows mark the phase transitions (compare
with Fig.~\ref{fig:fig2}(b)). There are the obvious Van der Waals
loops at high densities and smooth transitions at low densities.
(c) The set of isotherms for $\sigma_1=1.55$. The arrows mark the
phase transitions (compare with Fig.~\ref{fig:fig2}(c)).}
\end{figure}

In Fig.~\ref{fig:fig3} we present the isotherms for
$\sigma_1=1.15, 1.35, 1.55$ at different temperatures. In
Fig.~\ref{fig:fig3}(a) there are only Van der Waals loops,
corresponding to the first-order liquid-triangular lattice
transition. The cases $\sigma_1=1.35, 1.55$ are much more complex.
Fig.~\ref{fig:fig3}(b) represents the low-temperature and
high-temperature sets of isotherms for $\sigma_1=1.35$. One can
see that at low temperatures there are four regions on the
isotherms corresponding to the phase transitions, the low density
ones being smooth as in the case of liquid-hexatic-solid
transition \cite{binder} and the high densities  part containing
the Van der Waals loops characteristic of the first order phase
transition. At high temperatures (see Fig.~\ref{fig:fig3}(c))
there is only one liquid-triangular lattice first-order
transition. From Fig.~\ref{fig:fig3}(b) one can guess that the
melting of the low-density and high-density parts of the phase
diagram occurs with different scenarios: at low densities the
KTHNY scenario is probable, while the high density phase melts
through the first-order phase transition. As we are going to show
in the following, the intermediate region between the low density
triangular solid and the (normal) fluid can be qualified as
hexatic phase. The case $\sigma_1=1.55$ (Fig.~\ref{fig:fig3}(c))
is similar to the previous one: the low density part is smooth as
in the case of liquid-hexatic-solid transition \cite{binder} and
the high densities  part demonstrates the Van der Waals loops. It
is possible to expect the continuous transition at low densities
and the first order transition at high densities.

\begin{figure}
\includegraphics[width=8cm]{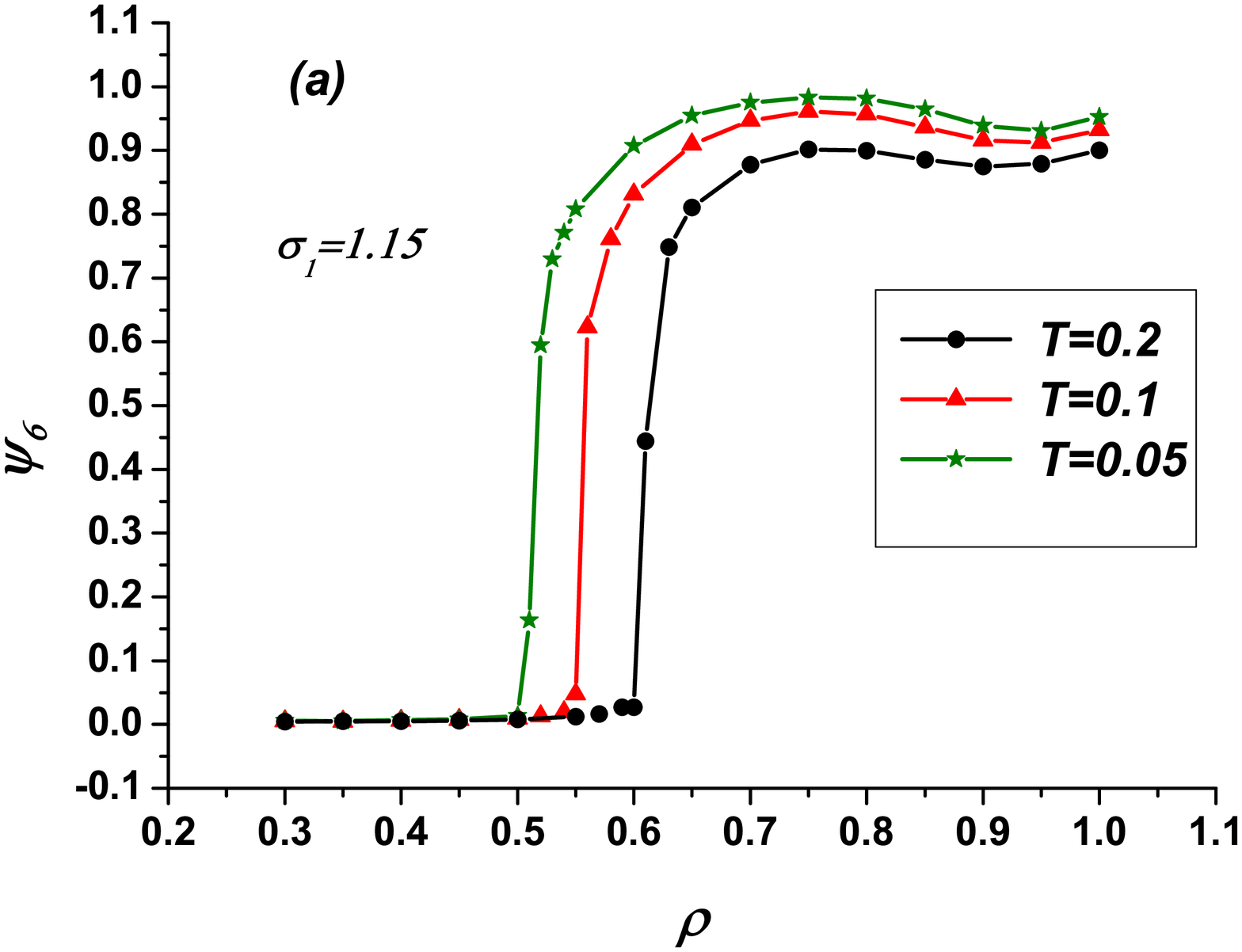}%

\includegraphics[width=8cm]{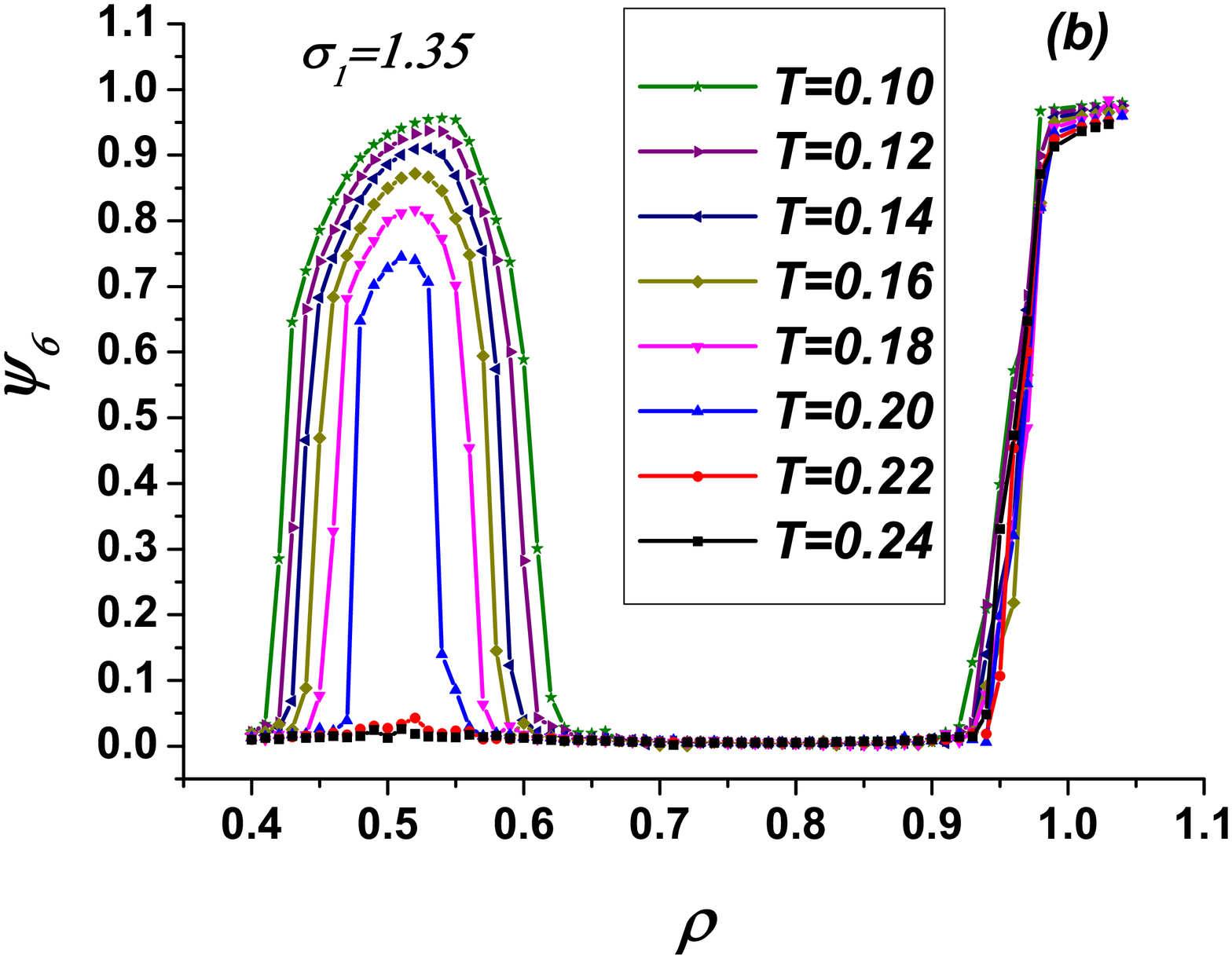}%

\includegraphics[width=8cm]{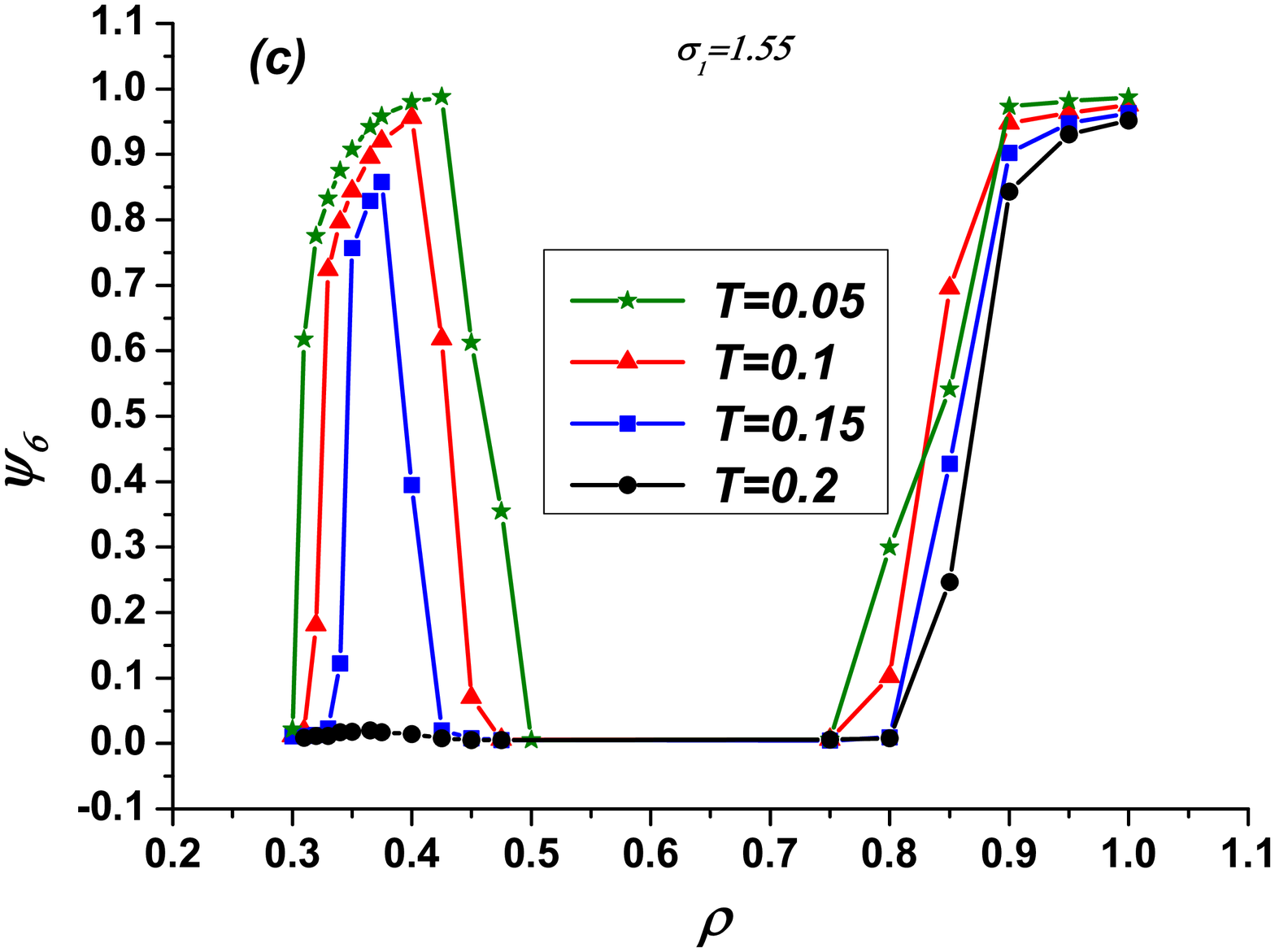}%

\caption{\label{fig:fig4} (Color online) Orientational order
parameter $\psi_6$ as a function of density for different
temperatures (a) $\sigma_1=1.15$; (b) $\sigma_1=1.35$; (c)
$\sigma_1=1.55$. }
\end{figure}

In Fig.~\ref{fig:fig4}, the orientational order parameter (OOP) is
presented as a function of density for a set of temperatures for
three values of the width of the repulsive shoulder
$\sigma_1=1.15; 1.35; 1.55$. For $\sigma_1=1.15$
(Fig.~\ref{fig:fig4}(a)) there is an abrupt quasilinear change of
OOP. We attribute this behavior to the weak first-order
liquid-triangular solid transition.

As it was mentioned above, the phase diagram for $\sigma_1=1.35$
is much more complex. One can see (Fig.~\ref{fig:fig4}(b)), the
behavior of OOP is different depending on the location on the
phase diagram: at the low density part of the phase diagram OOP
behaves smoothly while at high densities one can see that the OOP
increase almost linearly in the density. Fig.~\ref{fig:fig4}(b)
also suggests that the melting at low densities is continuous,
while at high densities melting transition is of the first order.
The similar behavior was found for the case $\sigma_1=1.55$.

\begin{figure}

\includegraphics[width=8cm]{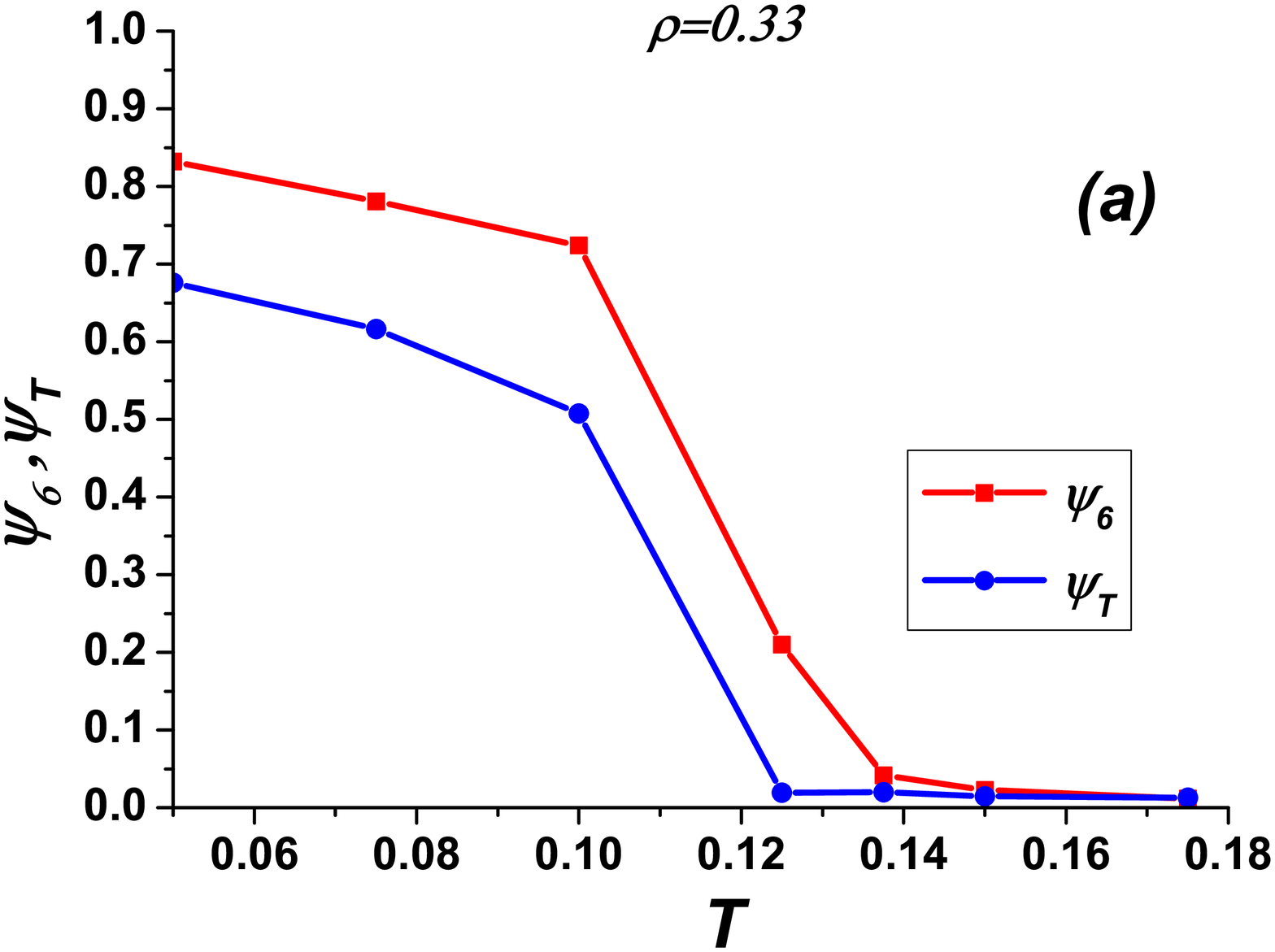}%

\includegraphics[width=8cm]{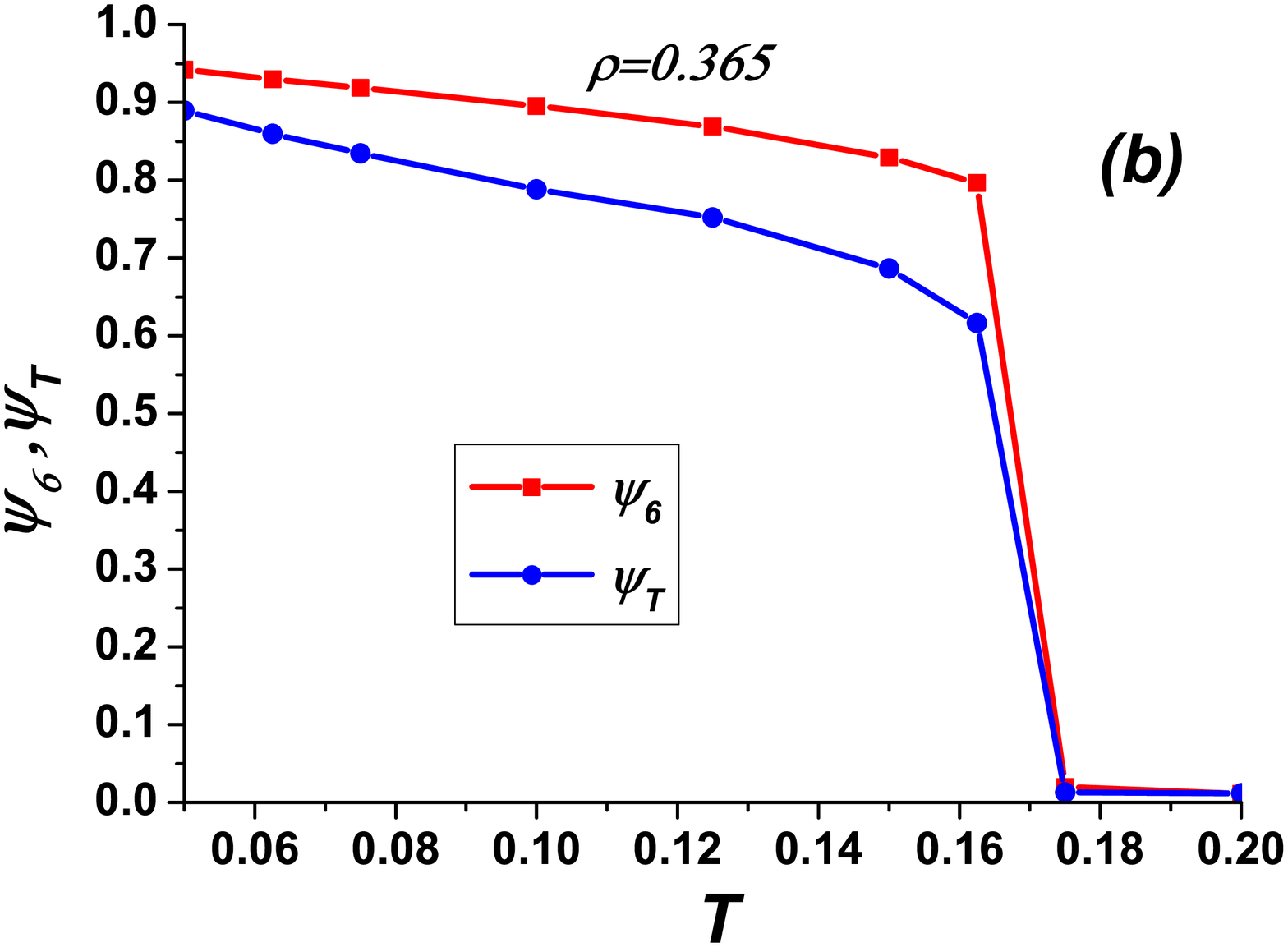}%

\includegraphics[width=8cm]{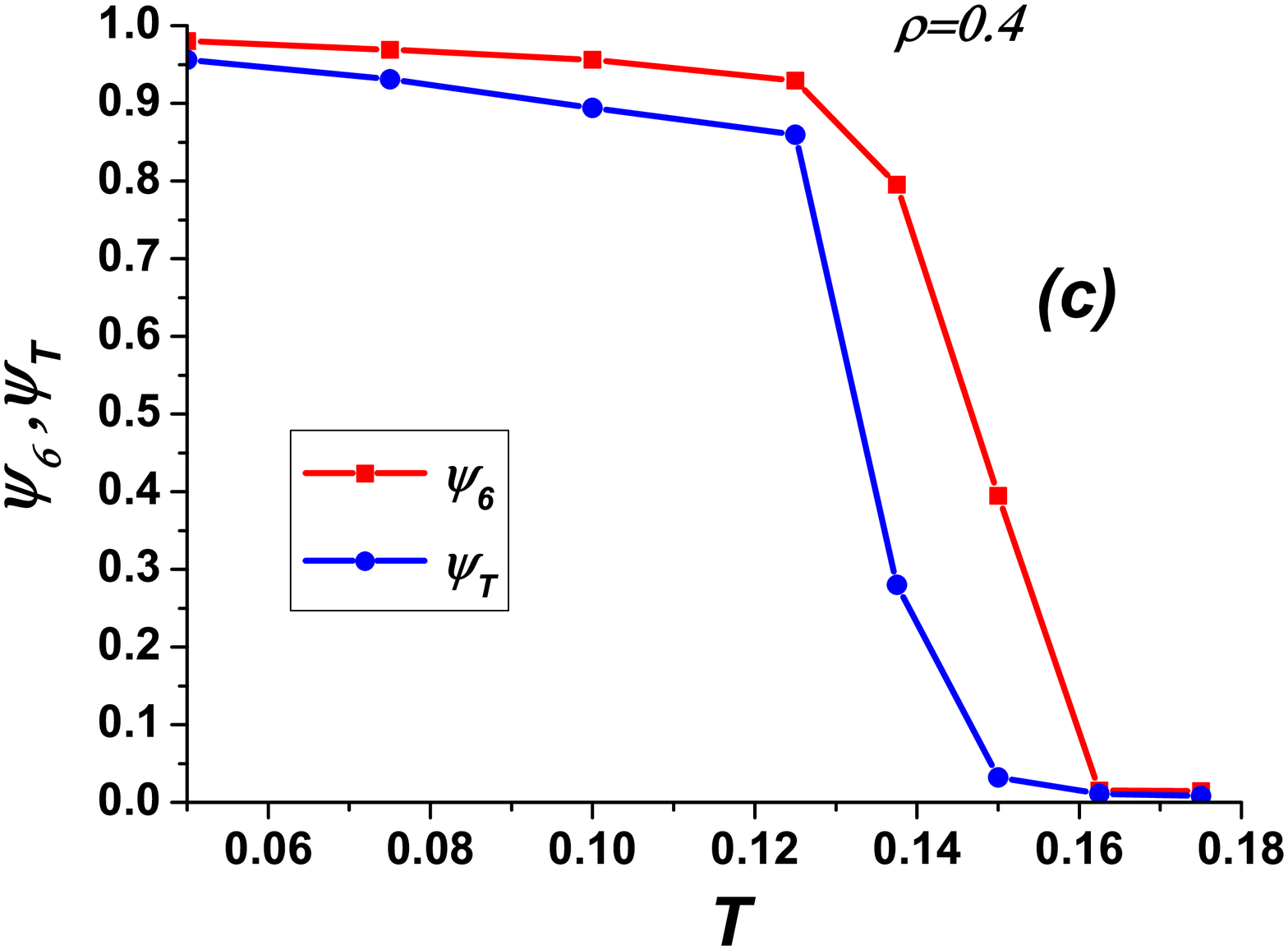}%

\caption{\label{fig:fig5} (Color online) OPs $\psi_T$ and $\psi_6$
as functions of temperature for $\rho=0.33$ (a); $\rho=0.365$ (b);
$\rho=0.4$ (c) in the case of $\sigma_1=1.55$. It is the narrow
hexatic phase in the cases (a) and (c), while it may be absent for
the case (b) (see discussion in the text).}
\end{figure}

In Fig.~\ref{fig:fig5}, we plot the translational and
orientational OPs for $\sigma_1=1.55$ and $\rho=0.33, 0.365, 0.4$
as a function of temperature (an analogous behavior was observed
for all the other densities). We see that $\psi_T$ vanishes at a
slightly smaller temperature than $\psi_6$, which implies that the
hexatic phase is confined to a narrow $T$ interval. It is
necessary to note, that in the case of the conventional
first-order phase transition, the density change at the melting
line maximum is equal to zero. For the density close to the
maximum on the low density part of the phase diagram
($\rho=0.48$), the region of the hexatic phase is very narrow,
however, we cannot conclusively determine whether the width of the
hexatic phase in the maximum point is equal to zero. In the case
of $\sigma_1=1.35$ the similar behavior takes place
\cite{dfrt1,dfrt2,dfrt3}.

\begin{figure}

\includegraphics[width=8cm]{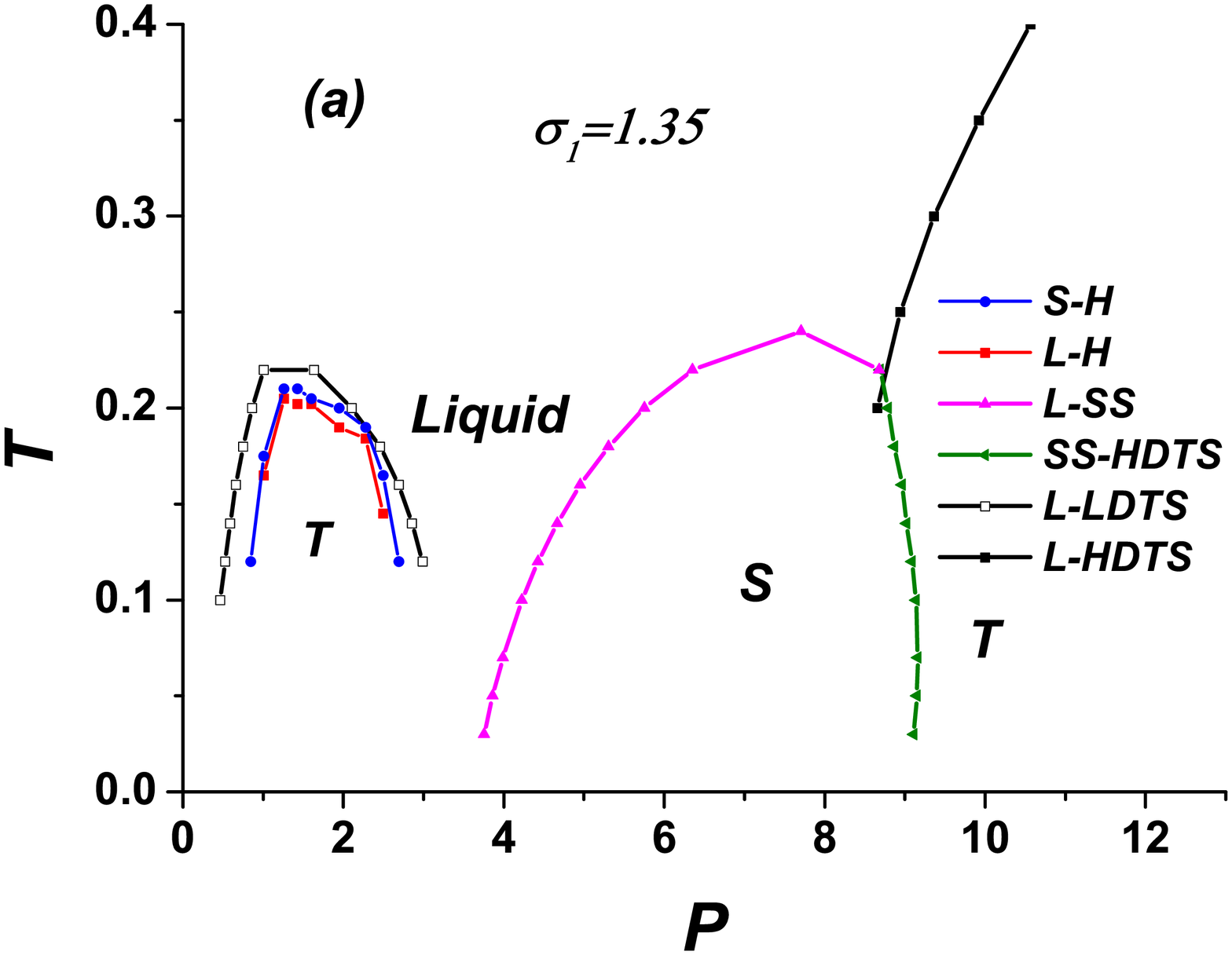}%

\includegraphics[width=8cm]{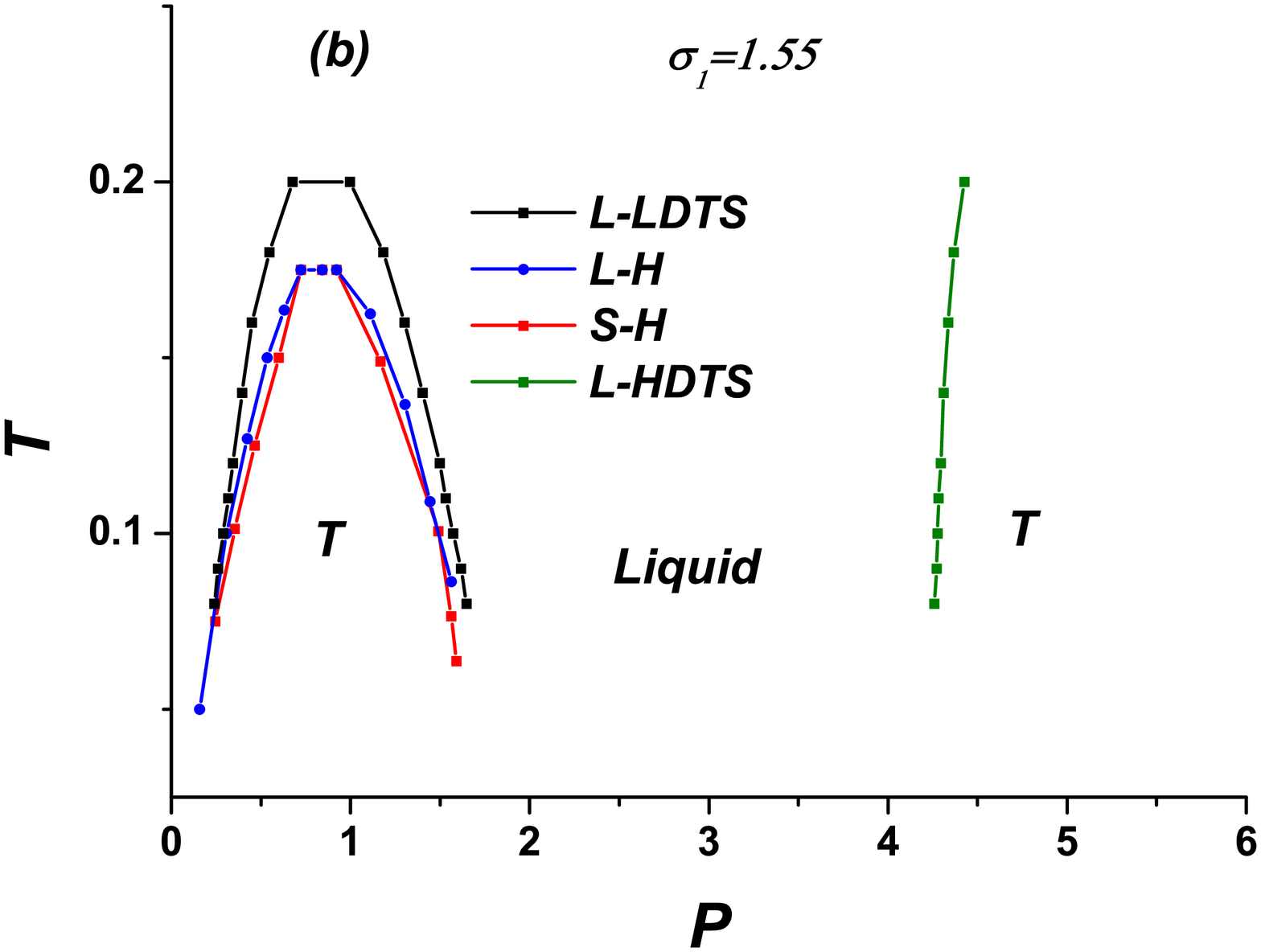}%

\caption{\label{fig:fig11} (Color online) The phase diagrams of
the systems with $\sigma_1=1.35$ (a) and $\sigma_1=1.55$ (b) in
$P-T$ plane (compare Fig.~\ref{fig:fig2}). There is a narrow
hexatic phase region in the low-density part of the phase diagram.
The lines of solid-hexatic and hexatic-liquid transitions were
obtained with the help of the calculations of $\psi_6$ and
$\psi_T$ (see Fig.~\ref{fig:fig5}). S-H means "solid-hexatic", L-H
- "liquid-hexatic", L-SS - "liquid-square lattice solid", SS-HDTS
- "square lattice solid - high density triangular lattice solid",
L-LDTS - "liquid - low density triangular lattice solid", L-HDTS -
"liquid - high density triangular lattice solid".}
\end{figure}

In Fig.~\ref{fig:fig11}, the phase transition lines of the
solid-hexatic and hexatic-liquid transitions, obtained from the
calculations of $\psi_6$ and $\psi_T$ (see, for example,
Fig.~\ref{fig:fig5}), are shown in comparison with the
solid-liquid transition lines (see Fig.~\ref{fig:fig2}) in $P-T$
planes for $\sigma_1=1.35$ and $\sigma_1=1.55$. One can see that
the transitions are mainly inside the solid region, obtained in
the framework of the free-energy calculations. This fact also
supports the idea that the melting in this region occurs through
two continuous transitions.

As it was mentioned above (see discussion after
Fig.~\ref{fig:fig2}(c)), one can expect that there are some
crystal phases inside the "gap"in the phase diagram
(Fig.~\ref{fig:fig11}(b)), which we could not find. In principle,
these phases can be obtained, for example, by the methods used in
\cite{prest5}. From Fig.~\ref{fig:fig3}(b) one can suppose that
these intermediate phases can exist at low enough temperatures. It
should be noted that the symmetry of these possible phases is
weaker than the triangular symmetry and favors the first-order
melting transition, because, as it was proposed in \cite{prest2}
for the square lattice, in this case the degree of the
orientational order is lower than in the triangular crystal.

The errors in calculation of the OOP $\psi_6$ are less than 1\%,
while the errors of the translational order parameter $\psi_T$ do
not exceed 5\%.

\begin{figure}

\includegraphics[width=6.5cm]{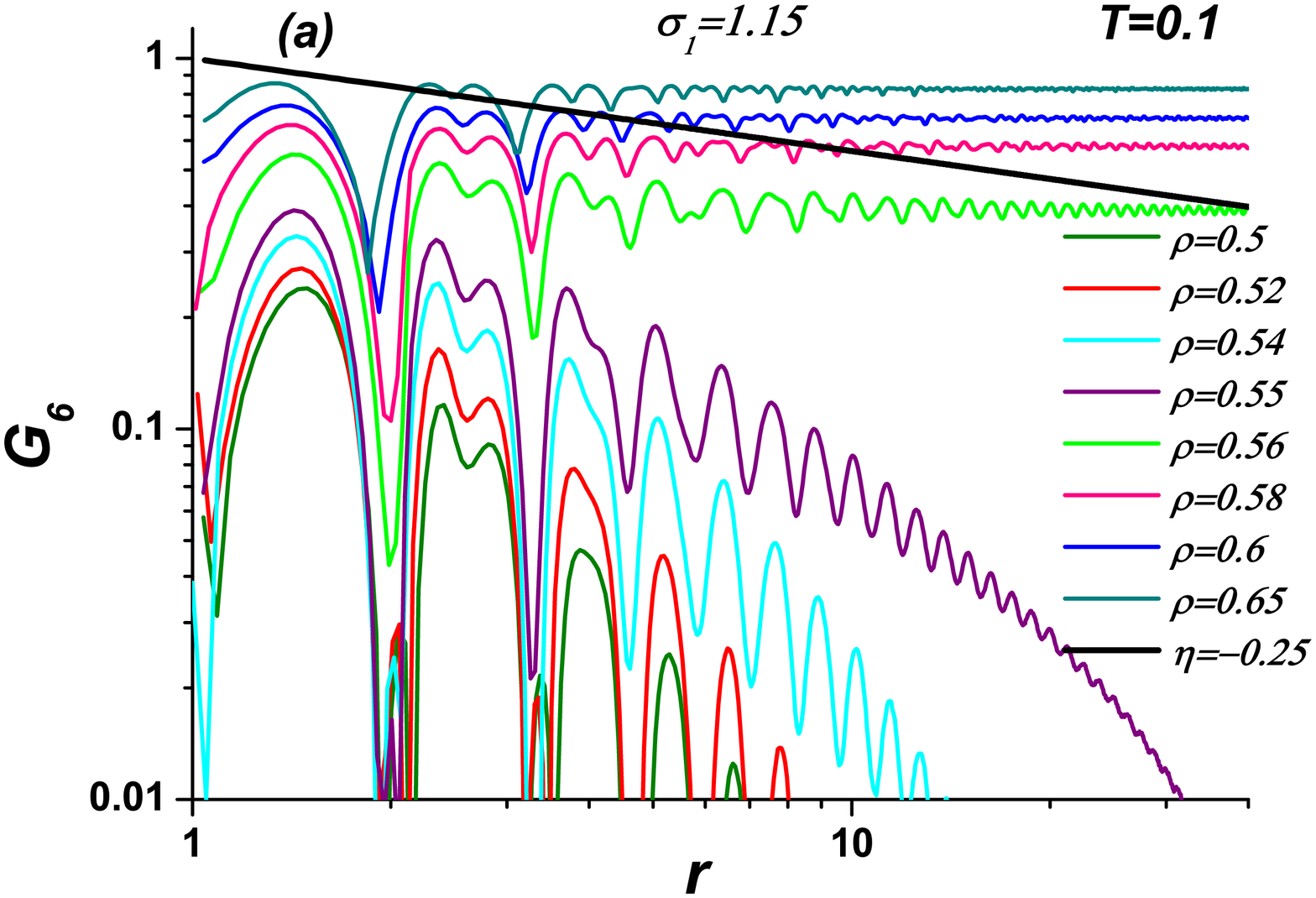}

\includegraphics[width=6.5cm]{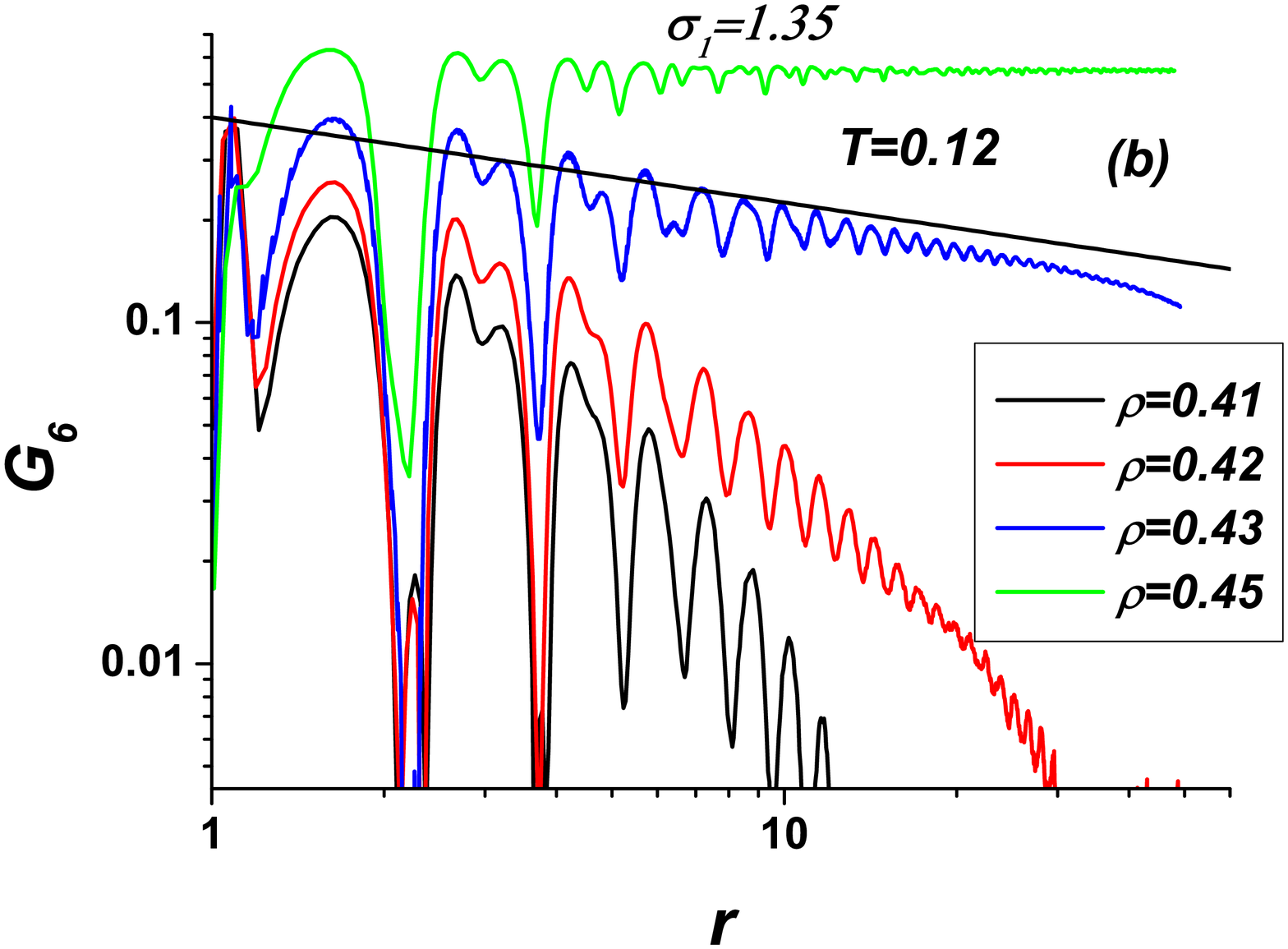}

\includegraphics[width=6.5cm]{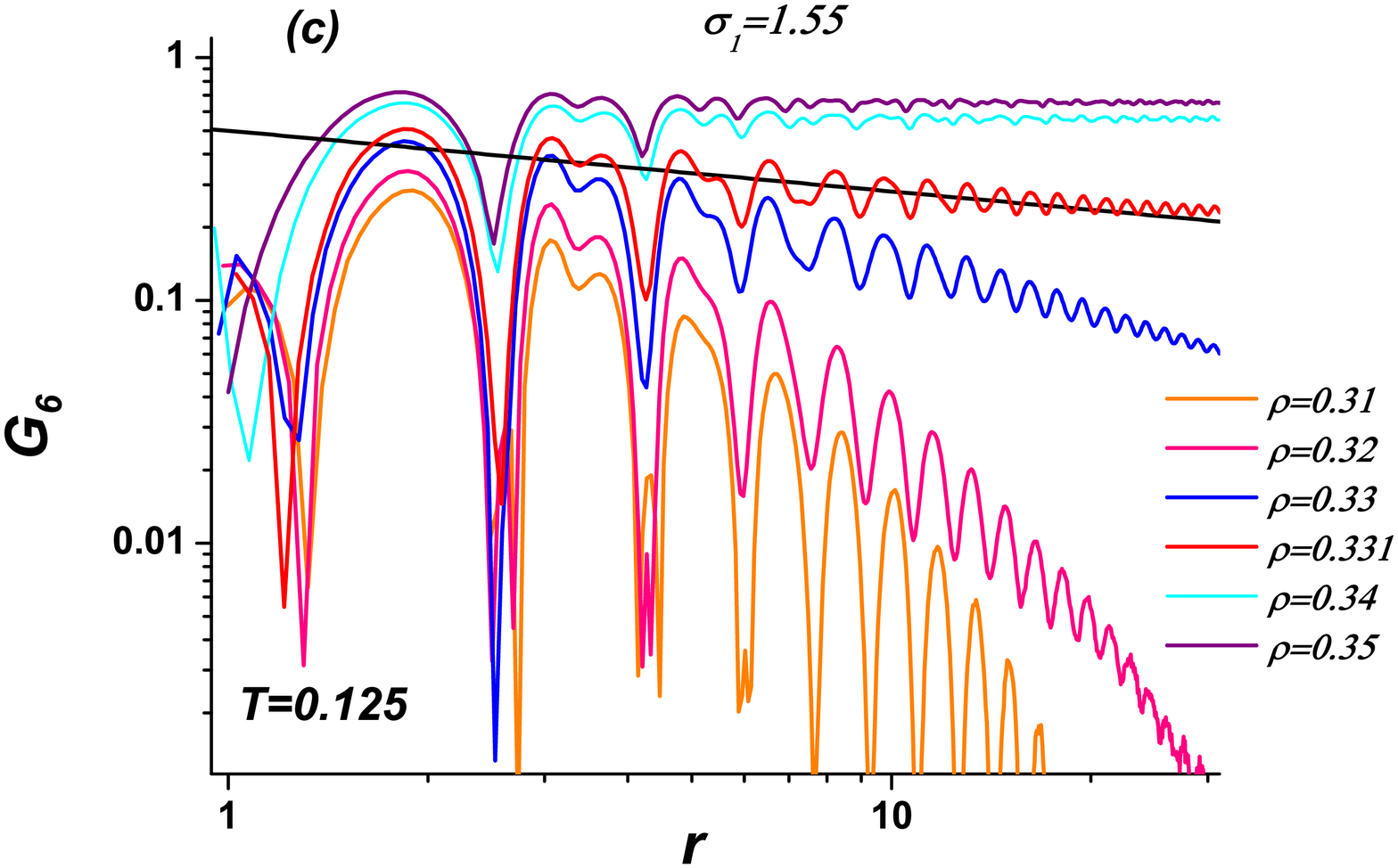}

\caption{\label{fig:fig7} (Color online) (a) Log-log plots of the
orientational correlation function $G_6(r)$ at selected densities
across the transition region for $\sigma_1=1.15$ and $T=0.1$. It
is seen that the behavior of $G_6(r)$ abruptly changes from
isotropic liquid like to the solid-like in accordance with the
first-order transition scenario. (b) Log-log plots of the
orientational correlation function $G_6(r)$ at selected densities
across the hexatic region for $\sigma_1=1.35$ and $T=0.12$. Upon
increasing $\rho$ from 0.41 to 0.45 there is a qualitative change
in the large-distance behavior of $G_6(r)$, from constant (solid)
to power-law decay (hexatic fluid), up to exponential decay
(normal fluid). Note that, consistently with the KTHNY theory, the
decay exponent $\eta$ is less than $1/4$ for $\rho>0.43$; (c) The
same as (b) for $\sigma_1=1.55$. The decay exponent $\eta$ is less
than $1/4$ for $\rho>0.33$. Straight lines correspond to the KTHNY
prediction $G_6(r)\propto r^{-1/4}$. }
\end{figure}

\begin{figure}

\includegraphics[width=8cm]{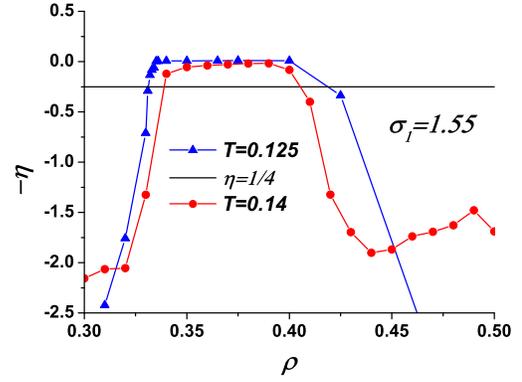}

\caption{\label{fig:fig13} (Color online)  Index $\eta$ as a
function of density for $\sigma_1=1.55$ and $T=0.125$ and
$T=0.14$. Horizontal line corresponds to $\eta=1/4$.}
\end{figure}

A more direct evidence of the hexatic phase emerges from the
large-distance behavior of OCF. Fig.~\ref{fig:fig7} shows these
functions for three values of the repulsive shoulder widths
$\sigma_1=1.15; 1.35; 1.55$ across the melting transition at
selected densities and temperatures. Calculations of the
orientational correlation function are made for 20000-102400
particles. From Fig.~\ref{fig:fig7}(a) one can see that the
behavior of $G_6(r)$ abruptly changes from isotropic liquid-like
with the exponential decay at large distances to the solid-like,
where OOF tends to the constant for large $r$, in accordance with
the first-order transition scenario.

In Fig.~\ref{fig:fig7}(b) we plot OOF at various densities across
the hexatic phase for $\sigma_1=1.35$ and $T=0.12$. Upon
increasing $\rho$ from $0.41$ to $0.45$ there is a qualitative
change in the large-distance behavior of $G_6(r)$, from constant
(solid) to power-law decay (hexatic fluid), up to exponential
decay (normal fluid). Note that, consistently with the KTHNY
theory, the decay exponent $\eta$ is less than $1/4$ for
$\rho>0.43$.

Fig.~\ref{fig:fig7}(c) is qualitatively similar to the
Fig.~\ref{fig:fig7}(b). For example, at $T=0.125$ the decay
exponent $\eta$ is less than $1/4$ for $\rho>0.33$. Index $\eta$
is shown as a function of density $\rho$ for $T=0.125$ and
$T=0.14$ in Fig.~\ref{fig:fig13}. From Fig.~\ref{fig:fig13} one
can see that the points corresponding to the condition $\eta=1/4$
are consistent with the line of isotropic fluid-hexatic phase
transition at the phase diagram in Fig.~\ref{fig:fig11}(b). The
similar results were obtained in Ref.~\cite{dfrt3} for
$\sigma_1=1.35$. In principle, this approach can be applied for
the construction of the phase diagram, however, it is rather time
consuming and cannot give the possibility to calculate the line of
the solid-hexatic transition.

\begin{figure}

\includegraphics[width=8cm]{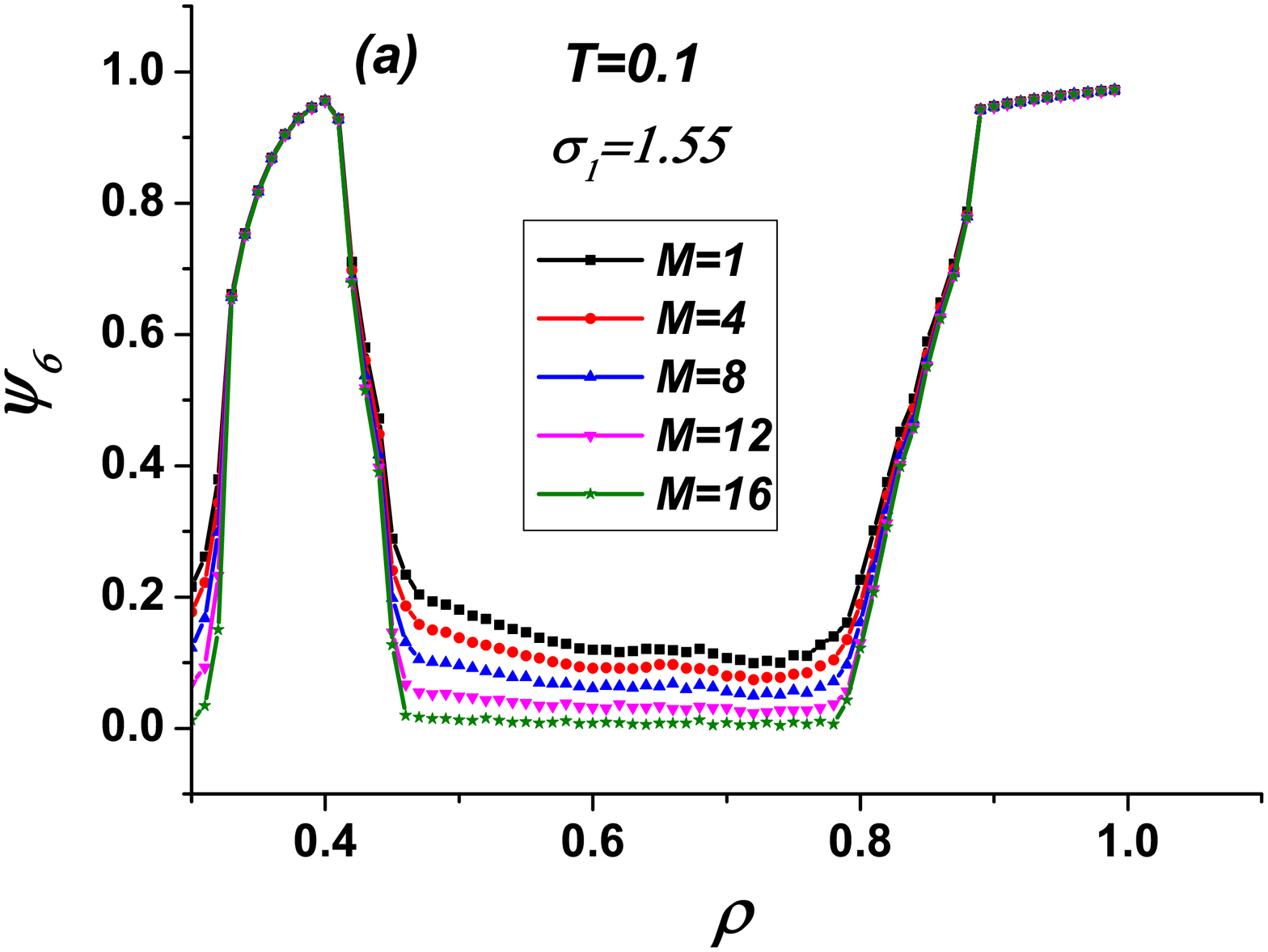}

\includegraphics[width=8cm]{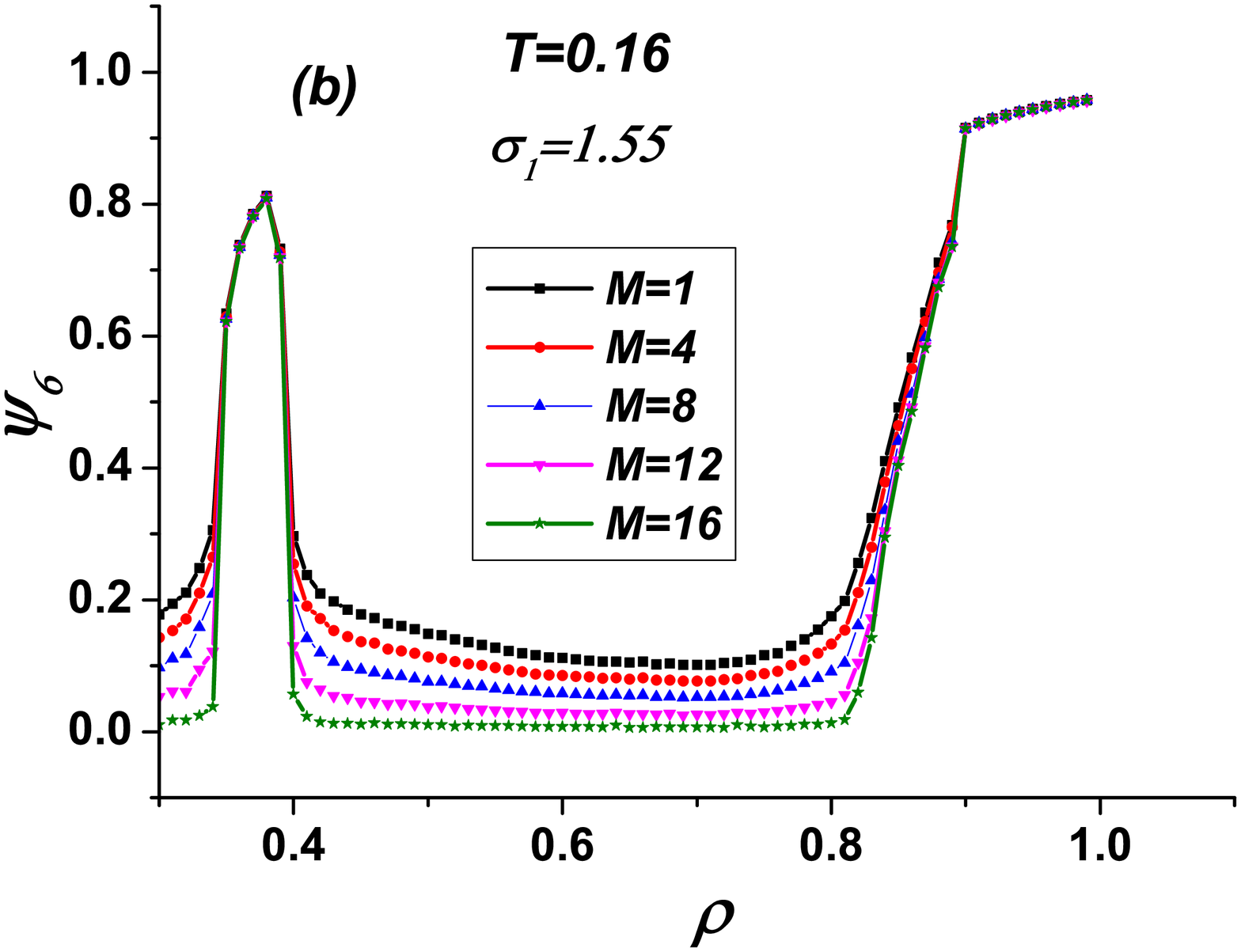}

\caption{\label{fig:fig9} (Color online) Bond orientational order
parameter $\psi_6$ for $\sigma_1=1.55$ and $T=0.1$ (a) and
$T=0.16$ (b) as a function of density $\rho$ for selected subbox
sizes $L = S/M$, for an $S\times S$ system with periodic boundary
conditions containing $N = 102400$ particles. Note that the shape
of the simulation box is a parallelepiped compatible with the
triangular lattice structure, and the parameter $M$ is equal to
the number of subboxes along the edge of the total system.}
\end{figure}

It should be noted, that the scaling analysis made in accordance
with the algorithm in Refs.~\cite{binderPRB,binder} also supports
the melting scenarios described above. For OOP we used a systems
of 102400 particles which were divided in subboxes. The subbox
size parameter $M$ is equal to the number of subboxes along the
edge of the total system and varies in our simulations from 1 to
16. As expected (see \cite{binderPRB,binder}), the
bond-orientational order parameter does not change in the ordered
region while it increases with increasing the number of the
subboxes in the liquid phase. An example of such behavior is shown
in Fig.~\ref{fig:fig9} for $\sigma_1=1.55$ and $T=0.1$ and
$T=0.16$.

In the case of $\sigma_1=1.35$ the similar analysis was made for
the melting of the square lattice region of the phase diagram, and
it was shown that the square lattice melts through the first-order
phase transition.

\section{Conclusions}

In conclusion, we have compared the phase diagrams of
two-dimensional ($2D$) classical particles repelling each other
through an isotropic core-softened potential for three values of
the soft core $\sigma_1=1.15; 1.35; 1.55$ corresponding to the
increasing softness of the potential from the almost vanishing to
rather large. For the smallest value of the soft core, the
potential of the system (\ref{3}) is close to the ordinary soft
sphere potential $U(r) \propto r^{-\frac{1}{14}}$. It is widely
believed that the $2D$ systems with hard potentials melt through
the weak first-order transition. We found that the system with
$\sigma_1=1.15$ follows this melting scenario in all range of the
thermodynamic parameters. On the other hand, one expects that the
$2D$ melting scenario is in accordance with the KTHNY one for the
softer potentials. For larger values of the soft core
$\sigma_1=1.35; 1.55$, the behavior of the system described by the
potential (\ref{3}) is determined by the soft long-range part of
the potential at low densities. The hard core of the potential
plays the main role at the high densities and temperatures. It
seems that this is the reason of the observed peculiarities of the
phase diagrams (see Fig.~\ref{fig:fig11}). We have provided the
evidences of the occurrence of two-stage continuous reentrant
melting via a hexatic phase in the $2D$ core-softened model at low
densities and validated a number of KTHNY predictions for this
part of the phase diagram. At the same time, at high densities the
melting occurs through the conventional first-order phase
transition.

These results may be useful for the qualitative understanding the
behavior of the systems with thermodynamic and dynamic anomalies
including the water monolayers confined between two hydrophobic
plates \cite{rev1,rice,barbosa,buld2d,krott,krott1}.

\section{Acknowledgments}

We are greateful to S. M. Stishov, V. V. Brazhkin, and O.B. Tsiok
for stimulating discussions and M.V. Kondrin for the help in
computer simulations. Yu.F. and E.T. also thank the Russian
Scientific Center Kurchatov Institute and Joint Supercomputing
Center of the Russian Academy of Science for computational
facilities.  The work was supported by the Russian Scientific
Foundation (Grant No 14-12-00820).


\end{document}